\documentclass[journal,12pt,onecolumn,draftclsnofoot]{IEEEtran}
\IEEEoverridecommandlockouts

\usepackage[table]{xcolor}  
\usepackage{xcolor}
\usepackage{multirow}
\usepackage[normalem]{ulem}
\usepackage[symbol]{footmisc}
\usepackage{amsthm}
\usepackage{pifont}
\usepackage[ruled,linesnumbered,vlined]{algorithm2e}
\usepackage{algorithmic}
\usepackage{makecell}

\usepackage{color,soul}

\usepackage{cite}
\usepackage{amsmath,amssymb,amsfonts}
\usepackage{graphicx}
\usepackage{textcomp}
\usepackage{mathtools}

\def\BibTeX{{\rm B\kern-.05em{\sc i\kern-.025em b}\kern-.08em
    T\kern-.1667em\lower.7ex\hbox{E}\kern-.125emX}}

\let\OldTexttt\texttt
\renewcommand{\texttt}[1]{\OldTexttt{\small{#1}}}

\usepackage{stackrel}

\begin{document}

\title{
RIC-O: Efficient placement of a disaggregated and distributed RAN Intelligent Controller with dynamic clustering of radio nodes
}

\author{Gabriel M. Almeida, Gustavo Z. Bruno, Alexandre Huff, \\ Matti Hiltunen, Elias P. Duarte Jr.,
Cristiano B. Both, Kleber V. Cardoso
\IEEEcompsocitemizethanks{
\IEEEcompsocthanksitem Gabriel M. Almeida and Kleber V. Cardoso are with the Universidade Federal de Goiás (UFG), Brazil. E-mail:  \{gabrielmatheus, kleber\}@inf.ufg.br
\IEEEcompsocthanksitem Gustavo Z. Bruno and Cristiano B. Both are with the University of Vale do Rio dos Sinos (UNISINOS), Brazil. E-mail: zanattabruno@edu.unisinos.br, cbboth@unisinos.br
\IEEEcompsocthanksitem Alexandre Huff is with Universidade Tecnológica Federal do Paraná, Brazil. E-mail:alexandrehuff@utfpr.edu.br
\IEEEcompsocthanksitem Matti Hiltunen is with AT\&T Labs Research, USA. E-mail: hiltunen@research.att.com
\IEEEcompsocthanksitem Elias P. Duarte Jr. is with Universidade Federal do Paraná, Brazil. E-mail: elias@inf.ufpr.br
}
}

This work has been submitted to the IEEE for possible publication. Copyright may be transferred without notice, after which this version may no longer be accessible.

{\let\newpage\relax\maketitle}


\begin{abstract}
The Radio Access Network (RAN) is the segment of cellular networks that provides wireless connectivity to end-users. O-RAN Alliance has been transforming the RAN industry by proposing open RAN specifications and the programmable Non-Real-Time and Near-Real-Time RAN Intelligent Controllers (Non-RT RIC and Near-RT RIC). Both RICs provide platforms for running applications called rApps and xApps, respectively, to optimize the behavior of the RAN. We investigate a disaggregation strategy of the Near-RT RIC so that its components meet stringent latency requirements while presenting a cost-effective solution. We propose the novel RIC Orchestrator (RIC-O) that optimizes the deployment of the Near-RT RIC components across the cloud-edge continuum. Edge computing nodes often present limited resources and are expensive compared to cloud computing. For example, in the O-RAN Signalling Storm Protection, Near-RT RIC is expected to support end-to-end control loop latencies as low as 10ms. Therefore, performance-critical components of Near-RT RIC and certain xApps should run at the edge while other components can run on the cloud. Furthermore, RIC-O employs an efficient strategy to react to sudden changes and re-deploy components dynamically. We evaluate our proposal through analytical modeling and real-world experiments in an extended Kubernetes deployment implementing RIC-O and disaggregated Near-RT RIC.

\end{abstract}

 
\section{Introduction}\label{sec:intro}
The Radio Access Network (RAN) is considered the most critical segment of a cellular network \cite{pana22-5g-survey}. A RAN comprises network nodes that provide direct wireless connection to end-users. It requires continuous innovations to improve the overall mobile network performance and user experience. Novel RAN technologies have included innovations related to wireless communication technologies (e.g., mmWave, massive MIMO, intelligent reflective surfaces, and THz communications). Other innovations are related to current networking technologies, particularly network softwarization, virtualization, programmability, and management (e.g., NFV, SDN, Network Slicing, and Zero touch network \& Service Management). Traditionally, the RAN accounts for most of the CAPEX and OPEX costs of a wireless mobile network. Network operators also have to face several hurdles, such as interoperability issues, resulting from the diversity of hardware from different vendors and the burden of continuously supporting new services and applications. In this context, the O-RAN Alliance~\cite{O-RAN-Alliance} has been transforming the RAN industry by proposing open, virtualized, fully interoperable, and intelligent mobile networks \cite{garcia-saavedra2021-oran}.

The O-RAN Alliance defines a series of specifications describing open interfaces to ensure multi-vendor interoperability among the main components that form the RAN. That ``openness'' has been carefully introduced to allow interoperability without hurting the intellectual property of the participating companies. Given the significant investments necessary to promote innovation of RAN technologies, this approach is essential to keep the interest of those companies in contributing to this vital area. Yet another advantage of an open RAN is that it decreases the barriers for newcomers that can contribute to solving specific problems and move the technology further, particularly in academia. On a side note, it is important to highlight that O-RAN specifications~\cite{oran-arch} are aligned with 3GPP standards~\cite{3gpp-ts-38.401}. Therefore, it is reasonable to assume that as the 3GPP introduces enhancements, they are promptly incorporated by the O-RAN Alliance.

The new RAN architecture~\cite{oran-arch} stands out as a key contribution among the O-RAN specifications. The architecture combines concepts from SDN and NFV and also takes into account cloud-native technologies while largely adopting Artificial Intelligence (AI) and Machine Learning (ML) technologies \cite{balasubramanian2021-ric}. It adopts SDN concepts, such as control and data planes separation and the possibility of having a remote RAN controller. The design of a RAN controller presents formidable challenges, as it must support the execution of tens to hundreds of RAN functions while featuring a large protocol stack. The O-RAN architecture splits the controller into two main building blocks: the Near-Real-Time RAN Intelligent Controller (Near-RT RIC) for time-sensitive operations and the Non-Real-Time RAN Intelligent Controller (Non-RT RIC) for operations that present fewer time restrictions. Moreover, the O-RAN specifications standardize open interfaces among the architecture components, which run as virtual (network) functions (or services) on cloud-native infrastructures.

Those RAN controllers run AI/ML-based applications that establish control loops with the RAN nodes under their management. The Non-RT RIC runs applications called rApps that demand control loops to time intervals above 1s. The Near-RT RIC runs applications called xApps that establish control loops constrained to time intervals between 10ms and 1s. The time constraint of a given control loop depends on the RAN function under the management of the corresponding xApp. For example, an xApp related to medium access management may need to complete the control loop under the 10ms threshold, while an xApp related to user session management may tolerate longer delays of up to 1s. In a large RAN, the Near-RT RIC (or some of its components) and latency-sensitive xApps must be replicated and assigned to manage a limited set of RAN nodes, i.e., a cluster of RAN nodes. Determining the minimum number of Near-RT RIC instances and where they must run is a non-trivial resource allocation problem. The problem becomes even more challenging considering the dynamics of a mobile wireless network. While this problem has been previously identified~\cite{rimedolabs2022-near-rt-ric-arch}, some works rely on multiple instances of the Near-RT RIC~\cite{Salvo22}. Also, to the best of our knowledge, there is no comprehensive solution to the problem as we describe in the present work.

In this work, we propose a disaggregation strategy of the Near-RT RIC so that individual components can be distributed and placed across the cloud-edge continuum.
We also propose a RIC Orchestrator (RIC-O) to deploy (or place) and monitor the Near-RT RIC components so that they can meet the stringent latency requirements.
The RIC-O employs optimization to deploy the Near-RT RIC components across the cloud-edge continuum while keeping the overall cost as low as possible.
Edge computing nodes often present limited resources and are expensive compared to cloud computing nodes. While performance-critical components of the Near-RT RIC platform and certain xApps should run at the edge, other components can run on cloud nodes. Furthermore, RIC-O employs a fast and efficient strategy to react dynamically to sudden changes and redeploy components. 
We also explore the flexibility of the O-RAN architecture to introduce a proposal that replicates some specific components of the Near-RT RIC. We evaluate the proposal through analytical modeling and real-world experiments in an extended Kubernetes deployment that runs the RIC-O and the disaggregated Near-RT RIC.

The main contributions of this work can be summarized as follows:
\begin{itemize}
    \item Proposes a disaggregation strategy to place the Near-RT RIC components across the cloud-edge continuum;
    \item Formalizes the problem of minimizing the overall cost of the placement of Near-RT RIC components while ensuring the latency-sensitive control loop;
    \item Proposes a hybrid approach that combines heuristic and optimal strategies to quickly provide cost-efficient solutions for placing Near-RT RIC components;
    \item Provides performance evaluation results to illustrate the advantage of the proposed approach;
    \item All implementations are publicly available, including the source code\footnote[1]{https://github.com/LABORA-INF-UFG/paper-GGAMECK-2023}, thus, allowing the work reproducibility.
\end{itemize}

Section~\ref{sec:background} presents an overview of the O-RAN architecture, particularly of the Near-RT RIC. Section~\ref{sec:related_work} describes the related work. Section~\ref{sec:model} provides the system model, problem formulation, and the optimal and heuristic strategies for the placement of the Near-RT RIC components. Section~\ref{sec:ric_orchestrator} introduces the RIC orchestrator, i.e., the proposed architecture that allows efficient placement of Near-RT RIC components and clustering of RAN nodes across the cloud-edge continuum. The implementation of the RIC orchestrator is described in Section~\ref{sec:prototype}. Real-world experiments and an analytical performance evaluation are presented in Section~\ref{sec:eval}. Finally, Section~\ref{sec:conclusion} concludes the article and discusses future work.
\section{Background}\label{sec:background}

\begin{figure}[htb]
\centering
    \includegraphics[width=0.7\textwidth]{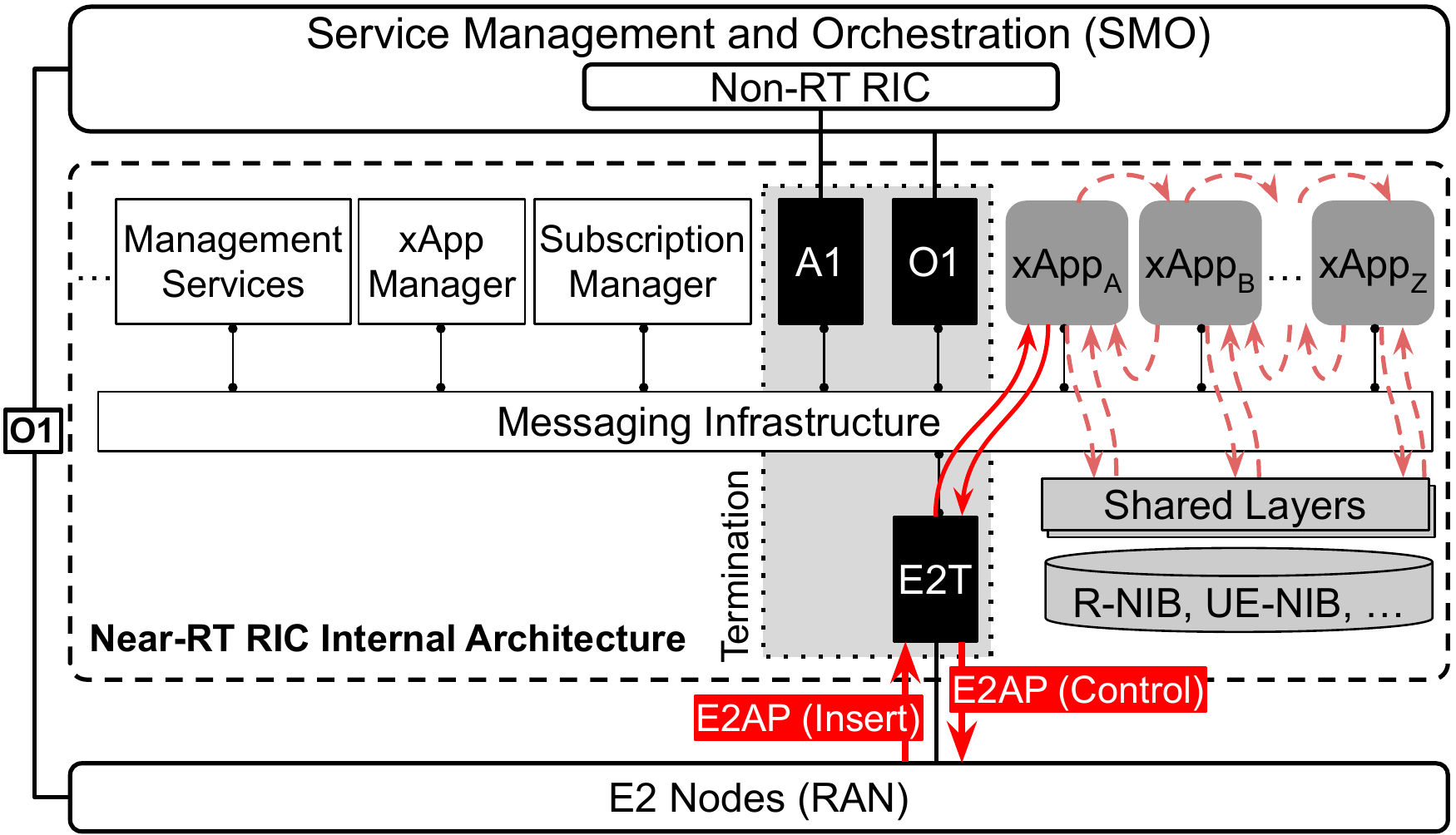}
\caption{O-RAN architecture.}
\label{fig:O-RAN_arch}
\end{figure}

The O-RAN architecture defines two RICs responsible for controlling and managing RAN nodes on different time scales. While the Non-RT RIC is in charge of tasks that can present a latency above 1s, the Near-RT RIC is responsible for tasks that demand a latency between 10ms and 1s. Figure~\ref{fig:O-RAN_arch} illustrates the O-RAN architecture highlighting the Near-RT RIC internal components. 
Several of those components are not latency-sensitive, such as the \texttt{Management Services}, the \texttt{xApp Manager}, the \texttt{Subscription Manager}, and \texttt{A1} and \texttt{O1} terminations.
However, \texttt{E2 Nodes}, E2 termination (\texttt{E2T}), \texttt{xApps}, and \texttt{Shared Layers} can be sensitive to latency and require correct placement.
The \texttt{xApp Manager} provides a flexible way for deploying and managing the near real-time applications (xApps).
 The xApps are applications that run services responsible for improving the performance of the RAN. Each \texttt{xApp} runs a well-defined function to access, control, and monitor \texttt{E2 Nodes}.
While some xApps may implement latency-sensitive functions, other xApps can also tolerate latencies up to 1s depending on the tasks they execute.
Moreover, xApps can provide services to other xApps. 
The \texttt{Subscription Manager} is responsible for providing E2 subscriptions from xApps to E2 nodes. The \texttt{Management Services} includes several services provided by the Near-RT RIC to the xApps and E2 nodes, such as routing management, alarm notification service, logging, and E2 node management. 

The \texttt{A1} termination also shown in Fig.~\ref{fig:O-RAN_arch} provides an interface between \texttt{Non-RT RIC} and \texttt{Near-RT RIC} to exchange information about policies and machine learning models that are executed by xApps. The \texttt{O1} termination represents the interface between the \texttt{Service Management and Orchestration} (\texttt{SMO}) framework with network functions and E2 nodes. This termination enables the \texttt{SMO} to execute Fault, Configuration, Accounting, Policy, and Security (FCAPS) operations on those network functions and E2 nodes. All O-RAN components are expected to support the \texttt{O1} termination when exchanging messages with \texttt{SMO}.
The \texttt{SMO} is responsible for managing and orchestrating the entire RAN, relying on the Non-RT RIC for RAN optimization, and the \texttt{O1} termination for RAN instrumentation.
Moreover, \texttt{SMO} is in charge of managing and orchestrating any cloud infrastructure in use\cite{oran-arch}.
The \texttt{E2T} is responsible for connecting the \texttt{Near-RT RIC} with one or more \texttt{E2 Nodes} using E2 Application Protocol (E2AP) and E2 Service Model (E2SM) which are described further. The \texttt{Messaging Infrastructure} element corresponds to the underlying messaging subsystem employed by all the Near-RT RIC components to exchange information \cite{O-RAN.WG3.E2GAP}. The \texttt{Shared Layers} correspond to the Shared Data Layer (SDL) and Shared Time-Series Layer (STSL), which are high-speed interfaces for accessing shared data storage by several stateless Near-RT RIC components and xApps. \texttt{R-NIB} and \texttt{UE-NIB} are examples of databases that store network information for RAN and User Equipment (UE), respectively.

According to the O-RAN specifications \cite{oran-arch}, RAN nodes (i.e., DUs, CUs, or O-RAN-compliant LTE eNBs) are \texttt{E2 Nodes} controlled through the E2 interface. This interface allows the Near-RT RIC to control procedures and functionalities of those nodes and is logically organized into two parts: E2AP and E2SM.
E2AP enables the communication between the Near-RT RIC and E2 nodes, which provide four services to the Near-RT RIC: Report, Insert, Control, and Policy \cite{O-RAN.WG3.E2GAP}.
The Report service allows xApps to subscribe to E2 nodes to receive information about specific RAN events. 
The Insert service also provides RAN information to xApps, but also allows xApps to configure E2 nodes through control messages.
The Control service allows the Near-RT RIC and xApps to send control messages to E2 nodes, which can initiate new procedures or resume a previously suspended procedure associated with that E2 node.
Such a procedure should specify exactly and completely the functional behavior of a given E2 node. Finally, the Near-RT RIC uses the Policy service to establish policy-driven monitoring and control the behavior of the corresponding E2 node \cite{O-RAN.WG3.E2AP}.
These services can be combined in different ways to implement E2SMs. An E2SM \cite{O-RAN.WG3.E2GAP} can be described as a contract between xApps and the RAN functions on E2 nodes.
Each RAN function exposed by a given E2SM allows the Near-RT RIC to monitor, suspend, stop, override or even control the behavior of the RAN on that E2 node.
Therefore, xApps and their corresponding E2 nodes must implement the same E2SM definitions to communicate and control the RAN. 

The E2AP Insert service is the most latency-sensitive loop of the Near-RT RIC. This loop starts when a given E2 node sends a message to the \texttt{E2T} using the E2AP (Insert) service. The \texttt{E2T}, in its turn, delivers that message to the corresponding \texttt{xApp}. This \texttt{xApp} then processes the message and sends back the corresponding reply using the E2AP Control service.
Moreover, the control loop can involve access to \texttt{Shared Layers} (i.e., SDL or STSL), it may also have to access one or more databases (e.g., \texttt{R-NIB} and \texttt{UE-NIB}) and even other \texttt{xApps}, as illustrated in Fig.~\ref{fig:O-RAN_arch}. We note that in some use cases, such as the O-RAN Signaling Storm Protection \cite{oran-use-cases}, it is essential to deliver the RIC control message back to the E2 node within the 10ms threshold. In this example use case, the E2 node suspends the current procedure execution and waits for the corresponding reply from an E2AP (Control) service to resume its operation. Although an E2 node can apply a default action when the 10ms threshold is exceeded, this usually implies applying further optimization procedures and performance penalties for the RAN.
In this work, we propose to disaggregate and distribute the Near-RT RIC components to guarantee the latency restrictions of the time-sensitive xApps, which execute under a control loop limit of 10ms. 


The O-RAN Software Community (OSC) provides an open-source reference implementation~\cite{oran-sc} of the Near-RT RIC. This implementation also functions as a baseline that allows the community to explore alternative deployments, including, for instance, replication, among other innovations. The SMO framework must be particularly aware of any alternative configuration so that it can dynamically (1) detect any latency violation (e.g., latency above the limit in the critical loop); (2) redeploy components to solve the violations; (3) re-orchestrate the Near-RT RIC components; and, (4) reconfigure all involved elements. As illustrated in Fig.~\ref{fig:O-RAN_arch}, the \texttt{SMO} communicates with all RAN elements, including the \texttt{E2 Nodes}, through the \texttt{O1} interface that offers the necessary means to reconfigure and reorganize the cluster of E2 nodes that each \texttt{E2T} and \texttt{xApps} are responsible for. Current SMOs do not fully implement all tasks in a coordinated manner. We claim one of the contributions of the current work is to present an SMO extension that can provide all necessary capabilities to accomplish those tasks coherently. Another contribution is related to task (2), which comprises the resource allocation problem, for which we formulate and propose optimal and heuristic strategies.

\section{Related Work}\label{sec:related_work}

This section presents an overview of related work that also has a focus on the RIC architecture. In a small RAN with a few dozen or less E2 nodes, it may be possible to deploy a single Near-RT RIC that satisfies the latency-sensitive control loop of the corresponding xApps. However, larger RANs with hundreds or even thousands of E2 nodes demand a different approach. Dryjanski and Kliks \cite{rimedolabs2022-near-rt-ric-arch} present two options for implementing the Near-RT RIC: centralized and distributed. In the centralized option, every E2 node (i.e., the whole gNB or eNB) is handled by the same and only Near-RT RIC, which can take unified decisions for an individual E2 node and globally optimize operations. In the distributed option, each E2 node type (i.e., O-CU, O-DU, or O-eNB) is handled by a specialized instance of the Near-RT RIC that allows optimizing these individual types of E2 nodes. The authors discuss the impact of these two implementation options in the design of the E2 interface. However, they do not tackle the distribution and replication of the Near-RT RIC components.

Singh and Nguyen \cite{Kumar22} propose a framework called O-RANFed to deploy and optimize a set of Federated Learning (FL) tasks that provide 5G slicing services in the context of the O-RAN specifications. To be precise: the authors introduce a theoretical model of the RIC architecture with support for FL. Moreover, the authors present an optimization model for the problem of local learning selection and resource allocation. The performance of FL improves with modeling and training done in every iteration. However, some details of the proposed FL-supporting RIC architecture are missing.

Huff, Hiltunen, and Duarte \cite{Huff21} discuss and evaluate techniques to make the RIC fault-tolerant while preserving high scalability. The fundamental assumption of this work is that traditional replication techniques cannot sustain high throughput and low latency as required by RAN elements. The authors propose techniques that use state partitioning, partial replication, and fast re-route with role awareness to decrease the overhead. Moreover, the authors implemented the fault tolerance techniques as a library called RFT (RIC Fault Tolerance) considering a distributed RIC, but do not deal with the problem of the placement of RIC components in a disaggregated virtual infrastructure.

D’Oro et al. \cite{Salvo22} introduced a novel orchestration framework called OrchestRAN for the Non-RT RIC. OrchestRAN allows network operators to specify high-level control and inference objectives. The orchestrator defines the optimal set of data-driven algorithms and their execution locations to achieve intents specified by the network operators. The work assumes that a complete instance of the Near-RT RIC is deployed to control each cluster of E2 nodes. Moreover, the authors show that the intelligence orchestration problem in O-RAN is NP-hard and design low-complexity approaches to support real-world applications. A prototype of OrchestRAN was implemented and tested at scale on Colosseum, i.e., the world’s largest wireless network emulator with hardware in the loop.

Schimidt, Irazabal, and Nikaein \cite{Schmidt21} presented FlexRIC, a software development kit to build specialized service-oriented controllers. FlexRIC has a centralized modular architecture with a minimal footprint and is designed for extensibility. The authors validate FlexRIC building implementations for two use cases: (i) a recursive RAN controller to allow multiple tenants to concurrently control and operate their services in a shared infrastructure over heterogeneous 5G networks, and (ii) an SD-RAN controller supplying programmability for RAN slicing with multi-radio technology, and flow-based traffic control aiming at low-latency communications.

Balasubramanian et al. \cite{Bharath21} disaggregate the traditional monolithic control plane RAN architecture. The authors introduce a Near-RT RIC platform that decouples the control and data planes of the RAN. The motivation of the project is to work towards intelligent and continuously evolving radio networks by fostering network openness and empowering network intelligence with AI-enabled applications. The authors provide functional and software architecture of the Near-RT RIC and discuss related design challenges. Moreover, they elaborate on how the Near-RT RIC can enable network optimization in 5G for the dual-connectivity use case using machine learning control loops. In this context, the Near-RT RIC architecture design is generic, providing several options for its implementation and deployment.

Other works focus on developing xApps and rApps. For example, Cao et al. \cite{Cao21}\cite{Cao22} propose an intelligent user access control scheme with Deep Reinforcement Learning (DRL). A federated DRL-based method is proposed with a global model server running on RIC that updates the distributed deep Q-networks (DQNs) parameters to optimize the performance of DQNs trained by UE. Johnson, Maas, and Van Der Merwe \cite{Johnson21} introduce NexRAN, a use case of the Powder mobile and wireless research platform, allowing closed-loop control of an approach for O-RAN slicing. O-RAN slicing, in this case, is implemented in the srsRAN open-source mobility stack and is exposed through a custom service model to xApps, executing on a Near-RT RIC.

\section{System Model, Problem Formulation, \\ and the Proposed Strategy}\label{sec:model}

In the following, we describe the system model, formulate the problem, and describe the optimal and heuristic strategies proposed to solve the problem.

\subsection{System Model}
\label{subsec:system_model}

\begin{figure}[htb]
\centering
    \includegraphics[width=0.6\textwidth]{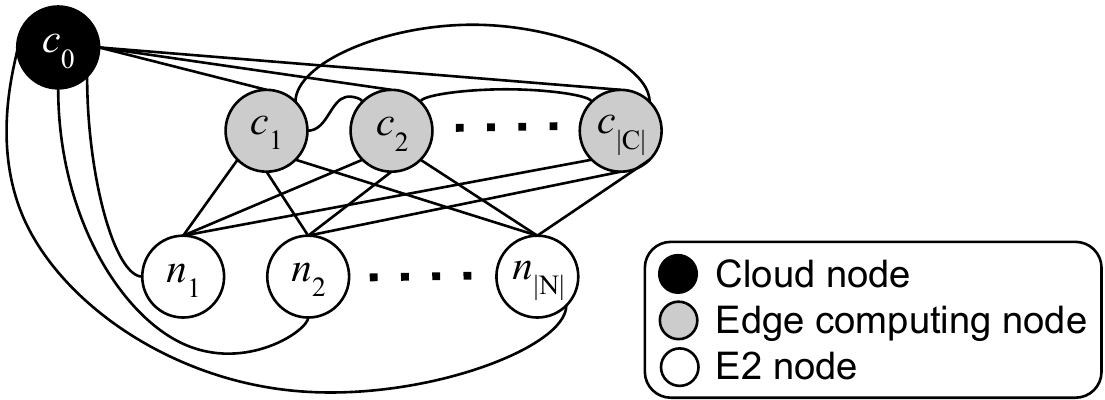}
\caption{Overlay communication network among E2 nodes and CNs.}
\label{fig:ran_topo}
\end{figure}

We assume a Radio Access Network that follows the O-RAN specifications and is composed of a set $N = \{n_1, n_2, ..., n_{|N|}\}$ of E2 nodes and a set $C = \{c_1, c_2, ..., c_{|C|}\}$ of edge computing nodes (CNs), where each CN is a host consisting of a general purpose hardware characterized by its processing capability $(c_m^{Proc})$, memory $(c_m^{Mem})$, and storage $(c_m^{Sto})$ resources. Moreover, we consider a node $c_0$ representing a cloud computing infrastructure (e.g., a large data center) with unlimited processing, memory, and storage resources but positioned far from the E2 nodes. We propose disaggregating the Near-RT RIC so that different components can run on various nodes. The node $c_0$ and each CN $c_m \in C$ can run any disaggregated RIC function as long as it does not exceed its resources. To represent the overlay communication network among all nodes, we define a graph $G = (\mathcal{V}, \mathcal{E})$ with $\mathcal{V} = \mathcal{V}_{C} \cup N$ representing the nodes (where $\mathcal{V}_{C} = \{c_0\} \cup C$), and $\mathcal{E} = \{e_{ij};v_i,v_j\in \mathcal{V}\}$ represent the set of links, where each link is characterized by its latency $e^{Lat}_{i, j}$. If $v_i=v_j$, then $ e^{Lat}_{i, j} = 0$, which is relevant for different components running on the same computing node.
As illustrated in Fig.~\ref{fig:ran_topo}, $c_0$ and each CN $c_m \in C$ can communicate with each other, while each E2 node $n_i \in N$ can communicate with $c_0$ or any CN $c_m \in C$. Therefore, the paths starting at E2 nodes and ending at computing nodes (that run RIC components) are defined over this graph $G$.

\begin{figure}[htb]
\centering
    \includegraphics[width=0.6\textwidth]{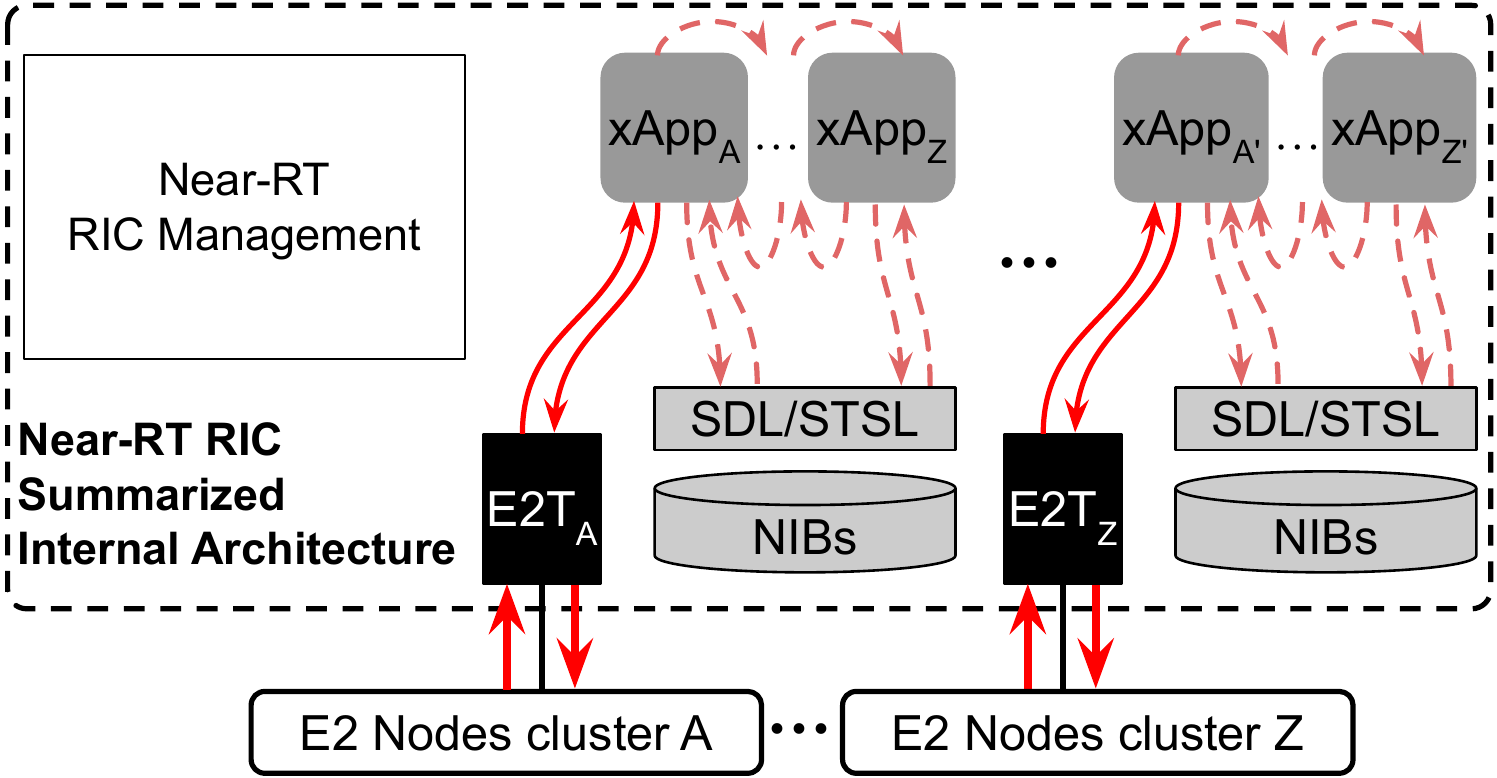}
\caption{Summarized version of the Near-RT RIC architecture.}
\label{fig:near-rt-ric_arch}
\end{figure}

We assume that the Near-RT RIC can be disaggregated and distributed into five main groups of components: Near-RT RIC Management (RIC\_Man), E2T, SDL/STSL, NIBs, and xApps, as illustrated in Fig.~\ref{fig:near-rt-ric_arch}. 
Each E2 node $n_i \in N$ is connected to an E2T component, which servers a set of xApps $A = \{\text{xApp}_1, \text{xApp}_2, ..., \text{xApp}_{|A|}\}$. 
While the control loop of a latency-sensitive xApp must be monitored per E2 node ($n_i \in N$), we are also interested in creating clusters of E2 nodes to minimize the replication of Near-RT RIC components, as illustrated in Fig.~\ref {fig:near-rt-ric_arch}. The red lines with arrows in this figure represent the control loop that each latency-sensitive xApp establishes with each E2 node. Given the computing nodes where the Near-RT RIC components are running and the graph from Fig.~\ref{fig:ran_topo}, it is possible to compute the round-trip latency starting from an E2 node $n_i \in N$, going through an E2T until an xApp (and possibly other RIC components), and back by the same path. 

For each E2 node $n_i \in N$, there are four atomic (or indivisible) groups: RIC\_Man, E2T, SDL/STSL, and NIBs, which the placement is represented by a tuple $p = (r, t, s, d)$.
We also employ the following auxiliary variables: $\mathcal{R}_{c_m} = \{0, 1\}$ that indicates if CN $c_m \in \mathcal{V}_{C}$ runs at least an instance of RIC\_Man, $\mathcal{T}_{c_m} = \{0, 1\}$ that indicates if CN $c_m \in \mathcal{V}_{C}$ runs at least an instance of E2T, $\mathcal{S}_{c_m} = \{0, 1\}$ that indicates if CN $c_m \in \mathcal{V}_{C}$ runs at least an instance of SDL/STSL, and $\mathcal{D}_{c_m} = \{0, 1\}$ that indicates if CN $c_m \in \mathcal{V}_{C}$ hosts at least one NIB instance. For example, the tuple $p=(c_0,c_1,c_1,c_1)$ represents RIC\_Man running on $c_0$ and the other atomic groups running on $c_1$, i.e., $\mathcal{R}_{c_0} = 1$, $\mathcal{T}_{c_1} = 1$, $\mathcal{S}_{c_1} = 1$, and $\mathcal{D}_{c_1} = 1$. Finally, we define a set of configurations $P = \{p_1, p_2, ..., p_{|P|}\}$ that lists all combinations of atomic groups running in the available computing nodes. However, the group of xApps can be further disaggregated and distributed into multiple computing nodes and, therefore, is not part of this configuration.

\subsection{Problem Formulation}
\label{subsec:prob_form}

The objective function minimizes the total cost to run a disaggregated and distributed Near-RT RIC. The cost has two parts, one fixed and another variable. The fixed cost corresponds to activating or leasing a CN, which is paid independently of the number of consumed resources. While the variable cost depends on the number of Near-RT RIC components running on a CN. Each computing node may be assigned a fixed cost, a variable cost, or both.

Let the decision variable $x^p_i \in \{0, 1\}$ represent the choice of configuration $p \in P$ to serve E2 node $n_i \in N$. Let $U_{c_m}^p = \{0, 1\}$ indicate if CN $c_m \in \mathcal{V}_{C}$ is used in configuration $p \in P$. The activation (or leasing) of CN $c_m \in \mathcal{V}_{C}$ is given by:
\begin{equation}
\label{eq:fixed_cost}
    \mathcal{U}_{c_m} = 
    \begin{cases}
        1,& \text{if} \sum\limits_{n_i \in N} \sum\limits_{p \in P} ( x^{p}_{i} \ U_{c_m}^{p} ) > 0\\
        0,& \text{otherwise}.
    \end{cases}
\end{equation}



The fixed cost is defined as:
\begin{equation}
\label{eq:phi1}
    \Phi_{Fix} =  \sum_{c_m \in \mathcal{V}_{C}} (\mathcal{U}_{c_m} F(c_m)),
\end{equation}
where $F(c_m) \in \mathbb{R}$ represents the cost to activate CN $c_m \in \mathcal{V}_{C}$. While the fixed cost may be less common in cloud-native infrastructures, it is included in the model to make it flexible enough to represent configurations such as bare-metal services.


Let the decision variable $y^a_{i,m} \in \{0, 1\}$ represent the choice of xApp $a \in A$, running on $c_m \in \mathcal{V}_{C}$, to control the E2 node $n_i \in N$. The utilization of CN $c_m \in \mathcal{V}_{C}$ to run xApp $a \in A$ is given  by:
\begin{equation}
\label{eq:xApps_cost}
    \mathcal{A}_{c_m}^a = 
    \begin{cases}
        1,& \text{if} \sum\limits_{n_i \in N} y^a_{i,m} > 0\\
        0,& \text{otherwise}.
    \end{cases}
\end{equation}


The variable cost is defined as:
\begin{align}
\label{eq:phi2}
    \Phi_{Var} = &\sum_{c_m \in \mathcal{V}_{C}} \Big( \mathcal{R}_{c_m} \mathcal{C}_{R}(c_m) + \mathcal{T}_{c_m} \mathcal{C}_{E}(c_m) + \mathcal{S}_{c_m} \mathcal{C}_{S}(c_m) \nonumber \\ 
    &+ \mathcal{D}_{c_m} \mathcal{C}_{D}(c_m) \Big) +
    \sum_{a \in A} \mathcal{A}_{c_m}^a \mathcal{C}_{A_a}(c_m),
\end{align}
where $\mathcal{C}_{R}(c_m)$, $\mathcal{C}_{E}(c_m)$, $\mathcal{C}_{S}(c_m)$, $\mathcal{C}_{D}(c_m)$, and $\mathcal{C}_{A_a}(c_m)$ represent the cost of running RIC\_Man, E2T, SDL/STSL, NIBs, and xApp $a \in A$, respectively, on a given CN $c_m \in \mathcal{V}_{C}$.

The objective function of minimizing the total cost is finally defined as:
\begin{equation}
    \textit{minimize} \ \ \ \ \ \Phi_{Fix} + \Phi_{Var}.
    \label{eq:OF}
\end{equation}

Next, we present the constraints of the problem. For each E2 node $n_i \in N$, exactly a single configuration $p \in P$ must be selected:
\begin{equation}
    \label{eq:c1}
    \sum_{p \in P} x^{p}_{i} = 1, \ \ \ \ \ \ \forall n_i \in N.
\end{equation}

The latency-sensitive control loop has two mandatory segments. The latency of the first one is measured between E2 node $n_i \in N$ and CN $c_m \in \mathcal{V}_{C}$, which runs the corresponding E2T:
\begin{equation*}
    \mathcal{L}_{i}^{c_\mathcal{T}} = \sum_{c_m \in \mathcal{V}_{C}} (\delta(x^{p}_{i}) \mathcal{T}_{c_m} e_{n_i, c_m}^{Lat}),
\end{equation*}
where $\delta(x^{p}_{i}) = \{0, 1\}$ indicates if CN $c_m \in \mathcal{V}_{C}$ runs the E2T that controls the E2 node $n_i \in N$. 

Concerning the second segment of the latency-sensitive control loop, its latency is measured between CN $c_m \in \mathcal{V}_{C}$ that runs E2T (associated with the E2 node $n_i \in N$) and CN $c_{m'} \in \mathcal{V}_{C}$ that runs xApp $a \in A$ (responsible for controlling the same E2 node $n_i \in N$):
\begin{equation*}
    \mathcal{L}_{i}^{c_\mathcal{T}, c_\mathcal{A}} = \sum_{c_m, c_{m'} \in \mathcal{V}_{C}} (\delta(x^{p}_{i},y^a_{i,m'}) \mathcal{A}^a_{c_{m'}} e_{c_m, c_{m'}}^{Lat}),
\end{equation*}
where $\delta(x^{p}_{i},y^a_{i,m'}) = \{0, 1\}$ indicates if CNs $c_m,c_{m'} \in \mathcal{V}_{C}$ run, respectively, E2T and xApp $a \in A$ responsible for controlling the E2 node $n_i \in N$.

In a latency-sensitive control loop, however, an xApp $a \in A$ (running on CN $c_m \in \mathcal{V}_{C}$) may need to access an SDL/STSL (running on CN $c_{m'} \in \mathcal{V}_{C}$) to reach a NIB (running in CN $c_{m''} \in \mathcal{V}_{C}$). This communication has the following latency:
\begin{align*}
    \mathcal{L}_{i}^{c_\mathcal{A},c_\mathcal{S},c_\mathcal{D}} = \psi(a) \sum_{c_m, c_{m'}, c_{m''} \in \mathcal{V}_{C}} \big(\delta(y^a_{i,m},x^{p}_{i}) \mathcal{S}_{c_{m'}} e_{c_m,c_{m'}}^{Lat} \nonumber \\
     + \delta'(y^a_{i,m},x^{p}_{i})\mathcal{D}_{c_{m''}} e_{c_{m'}, c_{m''}}^{Lat}\big),
\end{align*}
where $\psi(a) = \{0, 1\}$ indicates if xApp $a \in A$ needs to access SDL/STSL and NIB; $\delta(y^{a}_{i,m},x^{p}_{i}) = \{0, 1\}$ indicates if CNs $c_m,c_{m'} \in \mathcal{V}_{C}$ run, respectively, xApp $a \in A$ and SDL/STSL; and $\delta'(y^{a}_{i,m},x^{p}_{i}) = \{0, 1\}$ indicates if CNs $c_m,c_{m''} \in \mathcal{V}_{C}$ run, respectively, xApp $a \in A$ and NIB. These components are related to E2 node $n_i \in N$. 

Additionally, in a latency-sensitive control loop, an xApp $a \in A$ (running on CN $c_m \in \mathcal{V}_{C}$) may need to interact with other xApp $a' \in A$ (running on CN $c_{m'} \in \mathcal{V}_{C}$), which generates the following latency:
\begin{equation*}
    \mathcal{L}_{i}^{c_\mathcal{A}, c_{\mathcal{A}'}} = \psi'(a)\sum_{c_m, c_{m'} \in \mathcal{V}_{C}} (\delta(y^{a}_{i,m},y^{a'}_{i,m'}) \mathcal{A}^{a'}_{c_{m'}} e_{c_m, c_{m'}}^{Lat}),
\end{equation*}
where $\psi'(a) = \{0, 1\}$ indicates if xApp $a \in A$ needs to communicate with another xApp, and $\delta(y^{a}_{i,m},y^a_{i,m'}) = \{0, 1\}$ indicates if CNs $c_m,c_{m'} \in \mathcal{V}_{C}$ run, respectively, xApps $a, a' \in A$, which control the E2 node $n_i \in N$. 

Considering the potential chain of xApps $A_{ch} \subseteq A$ involved in the control loop of a certain xApp $a \in A$, its total latency is the sum of the four previous equations and cannot exceed the threshold $\rho_a$: 
\begin{align}
\label{eq:loop}
    &\mathcal{L}_{i}^{c_\mathcal{T}} + 
    \mathcal{L}_{i}^{c_\mathcal{T}, c_\mathcal{A}} + 
    \mathcal{L}_{i}^{c_\mathcal{A},c_\mathcal{S},c_\mathcal{D}} + \\ \nonumber
    &\sum_{a, {a'} \in A_{ch}} \big(\mathcal{L}_{i}^{c_\mathcal{A}, c_{\mathcal{A}'}} + \mathcal{L}_{i}^{c_\mathcal{A'},c_\mathcal{S},c_\mathcal{D}}\big) \leq \rho_a, 
    \forall n_i \in N, \forall a \in A.
\end{align}

The Near-RT RIC components running on a CN $c_m \in \mathcal{V}_{C}$ must not exceed its processing capability $(c_m^{Proc})$, memory $(c_m^{Mem})$, and storage $(c_m^{Sto})$ resources:
\begin{align}
    &\mathcal{R}_{c_m} \mathcal{R}^{Proc} + \mathcal{T}_{c_m} \mathcal{T}^{Proc} +  \mathcal{S}_{c_m} \mathcal{S}^{Proc} + \mathcal{D}_{c_m} \mathcal{D}^{Proc} +  \nonumber \\ 
    &\sum_{a \in  A_{c_m} \subseteq A} (\mathcal{A}^a_{c_m} {\mathcal{A}}^{a,Proc}) \leq c_m^{Proc}, \qquad\quad \ \ \forall c_m \in \mathcal{V}_{C}, \label{eq:c3}\\
    &\mathcal{R}_{c_m} \mathcal{R}^{Mem} + \mathcal{T}_{c_m} \mathcal{T}^{Mem} +  \mathcal{S}_{c_m} \mathcal{S}^{Mem} + \mathcal{D}_{c_m} \mathcal{D}^{Mem} + \nonumber \\
    &\sum_{a \in  A_{c_m} \subseteq A} (\mathcal{A}^a_{c_m} {\mathcal{A}^{a,Mem}}) \leq c_m^{Mem}, \qquad\quad \ \ \forall c_m \in \mathcal{V}_{C}, \label{eq:c4}\\
    &\mathcal{R}_{c_m} \mathcal{R}^{Sto} + \mathcal{T}_{c_m} \mathcal{T}^{Sto} +  \mathcal{S}_{c_m} \mathcal{S}^{Sto} + \mathcal{D}_{c_m} \mathcal{D}^{Sto} + \nonumber \\
    &\sum_{a \in  A_{c_m} \subseteq A} (\mathcal{A}^a_{c_m} {\mathcal{A}}^{a,Sto}) \leq c_m^{Sto}, 
    \qquad\qquad \ \ \forall c_m \in \mathcal{V}_{C}, \label{eq:c5}
\end{align}
where $\mathcal{R}^{\bullet}, \mathcal{T}^{\bullet}, \mathcal{S}^{\bullet}, \mathcal{D}^{\bullet}$, and ${\mathcal{A}}^{a,\bullet}$ represent processing, memory, and storage demands for RIC\_Man, E2T, SDL/STSL, NIBs, and xApp $a \in A_{c_m} \subseteq A$, respectively.


As discussed in \cite{MIQP_complexity}, the problem represented by Equation~(\ref{eq:OF}) is NP-Complete because it corresponds to a mixed-integer quadratic programming (MIQP) decision. The proof shows that there is a polynomial-length certificate for yes-instances of the MIQP decision so that the MIQP decision is in NP. Moreover, the proof that the MIQP decision is NP-complete consists of providing a polynomial-time reduction of the problem for determining whether there is a cut of cardinality at least $k$ in a graph $G$ (which is NP-complete) to an instance of the MIQP decision.


\subsection{Heuristic Strategy}
\label{subseq:heuristic}

MIQP problems such as the one represent by Equation~(\ref{eq:OF}) can be optimally solved using traditional solvers, e.g., IBM ILOG CPLEX. However, this problem has a large number of decision variables that can be estimated as $(|N|\cdot|\mathcal{V}_{C}|)^{comp} + (|N|\cdot|\mathcal{V}_{C}|)^{|A|}$, where $comp$ represents the number of Near-RT RIC components. For example, a configuration with $|N| = 100$ E2 nodes, $comp = 4$, $|\mathcal{V}_{C}| = 5$ computing nodes, and $|A| = 5$ xApps involves more than $3 \cdot 10^{13}$ optimization variables. Therefore, due to the computational time and the number of resources required, which are extremely high, solving the problem optimally is impractical, even for small instances of the problem, as we present in Section~\ref{sec:eval}. 




To deal with large instances of the problem and quickly provide solutions for RIC-O reorchestration requests, the efficient heuristic strategy described as Algorithm~\ref{alg:heuristic_solution} is proposed. The heuristic starts with a feasible solution and then seeks to improve the cost efficiency through a greedy strategy. The heuristic takes as input the same parameters used in the optimal strategy and produces as output the placement of the Near-RT RIC components across CNs, including xApps. The output is represented by $Places[n_i,\text{`Near-RT RIC component'}]$, where $n_i \in N$ identifies the E2 node that is associated with a `Near-RT RIC component', which can be RIC\_Man ($\mathcal{R}_i$), E2T ($\mathcal{T}_i$), SDL/STSL ($\mathcal{S}_i$), NIBs ($\mathcal{D}_i$), or xApp $a_i \in A_i$. Moreover, let $(Z_i,{\preceq}) = \mathcal{T}_i \cup A_i \cup \mathcal{S}_i \cup \mathcal{D}_i$ be the ordered set of components involved in the latency-sensitive control loop, as described in Section~\ref{subsec:system_model} and illustrated in Fig.~\ref{fig:near-rt-ric_arch}. We present additional details in the following.


\begin{algorithm}
\label{alg:heuristic_solution}
\caption{Heuristic strategy}
\DontPrintSemicolon
\small
\SetKwInOut{Input}{Input}
\SetKwInOut{Output}{Output}
\Input{The same input of the optimal strategy}
\Output{Placement of the Near-RT RIC components}
\BlankLine
\BlankLine
\ForAll{$n_i \in N$}
{
    \ForAll{$\zeta_i \in (Z_i,{\preceq})$}
    {
        $Places[n_i,\zeta_i] \gets$ closestCN($n_i,\zeta_i$)\\
    }
}
\ForAll{$\hat{\zeta_i} \in \mathcal{R}_i \cup (Z_i,{\preceq})$}
{
    $CNcost[\hat{\zeta_i}] \gets$ sortByDecreasingCost($\hat{\zeta_i}$,$\mathcal{V}_C$)\\
}
\ForAll{$n_i \in N$}
{
    $Places[n_i,\mathcal{R}_i] \gets$ rePlace($\mathcal{R}_i,CNcost[\mathcal{R}_i]$)\\
    \ForAll{$\zeta_i \in (Z_i,{\preceq})$}
    {
        $Places[n_i,\zeta_i] \gets$ rePlace($\zeta_i,CNcost[\zeta_i]$)\\
    }
}
\Return $Places[\bullet,\bullet]$
\end{algorithm}

Initially, for each $n_i \in N$, the heuristic strategy seeks to place every Near-RT RIC component $\zeta_i \in (Z_i,\preceq)$ at the closest CN $c_m \in \mathcal{V}_C$ with enough available resources, as shown in Algorithm~\ref{alg:heuristic_solution} (lines 1--3). This corresponds to the best solution regarding latency, but it ignores the cost of CNs and causes unnecessary replication of Near-RT RIC components, e.g., E2T and xApps, which also increases cost. Next (lines 4, 5), in $CNcost[\hat{\zeta_i}]$, the algorithm sorts in decreasing order of cost the list of CNs that are available for running each Near-RT RIC component $\hat{\zeta_i} \in \mathcal{R}_i \cup (Z_i,{\preceq})$. RIC\_Man is included because it is a Near-RT RIC component, despite of not being involved in the latency-sensitive control loop. The last part of the algorithm (lines 6--9) focuses on improving cost efficiency. First (line 7), the RIC\_Man component is placed on the cheapest CN $n_i \in N$ that has enough available resources. RIC\_Man tends to be placed on $c_0$ since this tends to be the cheapest CN, and, generally, function \texttt{rePlace} determines that node has enough available resources. Moreover, a single RIC\_Man is probably able to serve all E2 nodes. Last (lines 8, 9), the algorithm seeks to find the cheapest CN $n_i \in N$ for each Near-RT RIC component $\zeta_i \in (Z_i,\preceq)$. Therefore, in addition to checking CN resource availability, function \texttt{rePlace} also checks compliance with the latency-sensitive control loop. These final steps of the heuristic also tend to reduce the number of instances of Near-RT RIC components because a single component may be able to ensure the latency is within the required threshold for multiple E2 nodes. 

The proposed heuristic strategy solves the problem in polynomial time and produces satisfactory results regarding the placement of the Near-RT RIC components, as we present in Section~\ref{sec:eval}. To be precise, the asymptotic complexity of the Algorithm~\ref{alg:heuristic_solution} is $\mathcal{O}((comp + |A|)^2 \cdot |N| \cdot |\mathcal{V}_{C}|)$, which represents the placement of all Near-RT RIC components that serve all E2 nodes, considering, in the worst case, all available CNs.
\section{RIC Orchestrator} \label{sec:ric_orchestrator}

\begin{figure}[!h]
    \centering
    \includegraphics[width=0.7\textwidth]{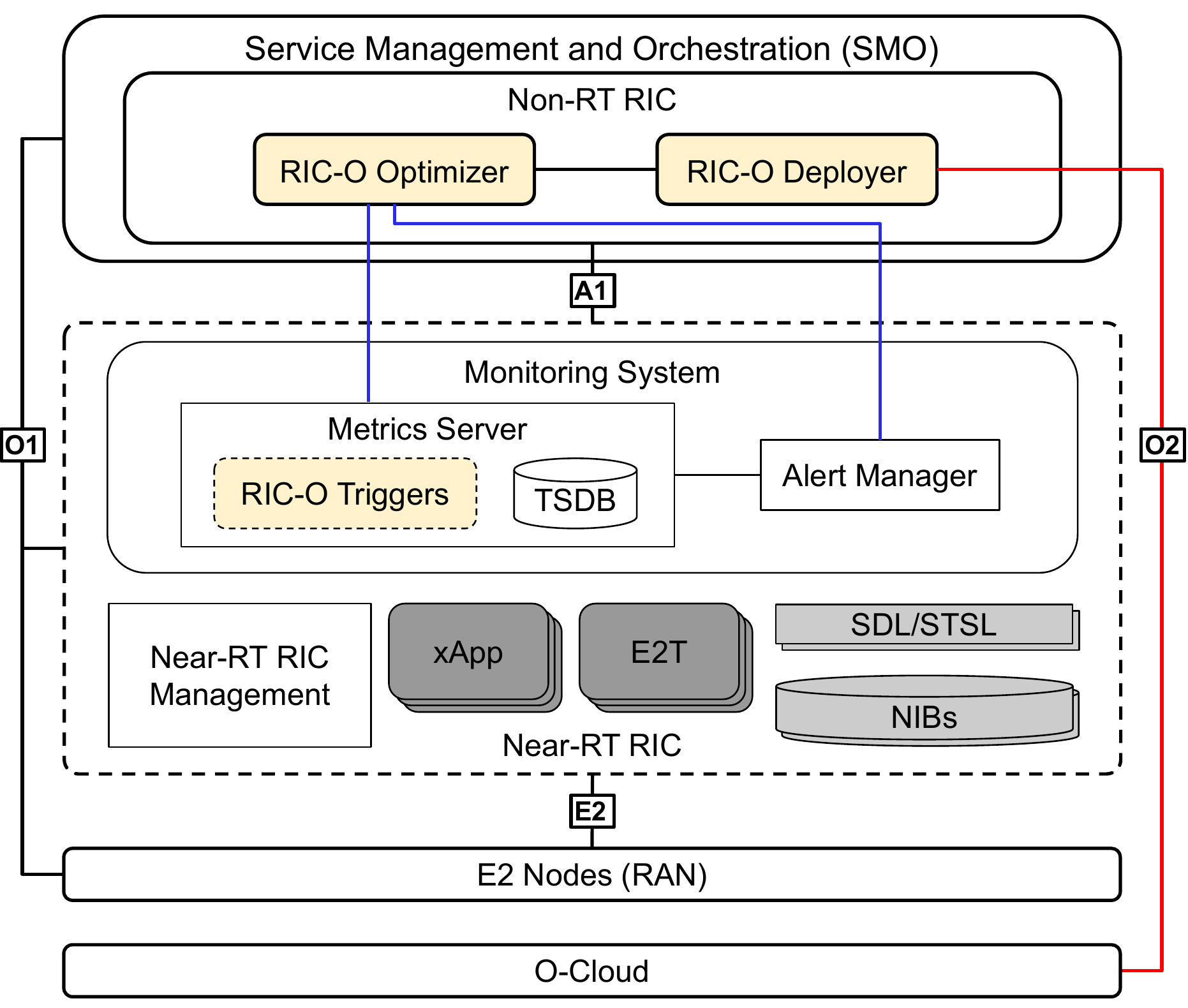}
    \caption{RIC-O architecture.}
    \label{fig:ric-o-arch}
\end{figure}

This section describes the architecture of the RIC-O which is responsible for the dynamic and efficient placement of the disaggregated Near-RT RIC components.
The architecture of the RIC-O is shown in Fig.~\ref{fig:ric-o-arch}, along with some of the RIC components required to describe the orchestrator. RIC-O was designed considering components that run in both Non-RT RIC and Near-RT RIC. The \texttt{RIC-O Optimizer} and \texttt{RIC-O Deployer} run on the Non-RT RIC, while the \texttt{RIC-O Triggers} execute on the \texttt{Metrics Server} of the Near-RT RIC. The \texttt{Monitoring System} is in charge of monitoring resource usage of the \texttt{Near-RT RIC} components, \texttt{E2 Nodes}, and \texttt{O-Cloud} infrastructure, in addition to throwing alerts whenever it detects threshold violations in the latency-sensitive control loop.

The \texttt{Metrics Server} is a monitoring module that collects metrics information from all \texttt{xApp} and \texttt{E2T} instances running on the \texttt{Near-RT RIC} platform. Example metrics include CPU utilization, memory usage, and the communication latency to monitored instances. The latency to access SDL/STSL shared layers and databases are also collected by the \texttt{Metrics Server}.
All metrics are stored in a time-series database (\texttt{TSDB}) for further processing and reporting.
The \texttt{RIC-O Triggers} component consists of predefined alert rules (e.g., the latency is above a given threshold) upon the occurrence of which an alert is generated. As the \texttt{Metrics Server} detects some metric violation that is associated with some trigger, it notifies the \texttt{Alert Manager} that a metric has been violated. The \texttt{Alert Manager}, in its turn, sends a message to the \texttt{RIC-O Optimizer} requesting a new round of optimization to reconfigure the system so that it can satisfy the metric that has been violated.

When a request is delivered to the \texttt{RIC-O Optimizer}, it fetches the current metric measurements from the \texttt{Metrics Server}, executing the optimization strategies described in Section~\ref{sec:model}. The heuristic and optimal optimization strategies run in parallel. The heuristic strategy should find a new solution quickly. However, the optimal strategy is executed even if the heuristic completes earlier. The solution generated by the optimal strategy is only applied if is superior to the heuristic and if it becomes available before a new notification of metric violation.
The \texttt{RIC-O Optimizer} sends the configuration of the computed solution to the \texttt{RIC-O Deployer}, which redeploys the \texttt{Near-RT RIC} components accordingly. The redeployment is only executed if the \texttt{RIC-O Optimizer} finds a solution different from the previous one. Lastly, the \texttt{RIC-O Deployer} uses the standardized \texttt{O2} interface defined by O-RAN to exchange messages with the \texttt{O-Cloud} element to effectively apply the new placement.

\section{Implementation} \label{sec:prototype}

We implemented a proof-of-concept prototype of RIC-O to validate and evaluate our proposal. The prototype is described in this section, along with the testbed used for its evaluation. \texttt{RIC-O Optimizer} and \texttt{RIC-O Deployer} were implemented using the Python language.
The \texttt{Monitoring System} component comprises the monitoring subsystem of the Near-RT RIC platform \cite{ric-alarm20} that runs on Kubernetes (K8S). The Near-RT RIC uses Prometheus to collect metrics, and \texttt{Metrics Server} corresponds to the main Prometheus server. Metrics are collected at the pod and cluster levels and stored as time series data in the internal Prometheus database (\texttt{TSDB}).
Prometheus was a natural choice to collect metrics in our proposal since it is the native solution of K8S for data collection and is part of the Near-RT RIC platform from the O-RAN SC.



We also employed the \texttt{Alert Manager} component from Prometheus to handle alerts issued by different client applications (e.g., \texttt{Metrics Server}). Alert handling may include deduplicating, grouping, silencing, inhibiting, and routing alerts to other endpoints. The \texttt{Alert Manager} notifies \texttt{RIC-O Optimizer} of the need for starting a new optimization round when a given RIC-O trigger detects a latency threshold has been violated.
The \texttt{RIC-O Triggers} element was implemented using the Prometheus Query Language (PromQL).
In this context, our prototype monitors CPU, memory, and network latency from K8S system pods, Near-RT RIC pods, and K8S nodes. One of the monitored metrics is the latency-sensitive control loop of the Near-RT RIC, which cannot surpass a given threshold (i.e., above 10ms for certain control loops). Upon reaching this threshold, a trigger launches the \texttt{RIC-O Optimizer} to compute a new placement solution for the Near-RT RIC components. After the optimization is computed, a notification is delivered to \texttt{RIC-O Deployer}, which redeploys the Near-RT RIC components based on the outcome of the \texttt{RIC-O Optimizer}.

\begin{figure}[!h]
    \centering
        \includegraphics[width=0.8\textwidth]{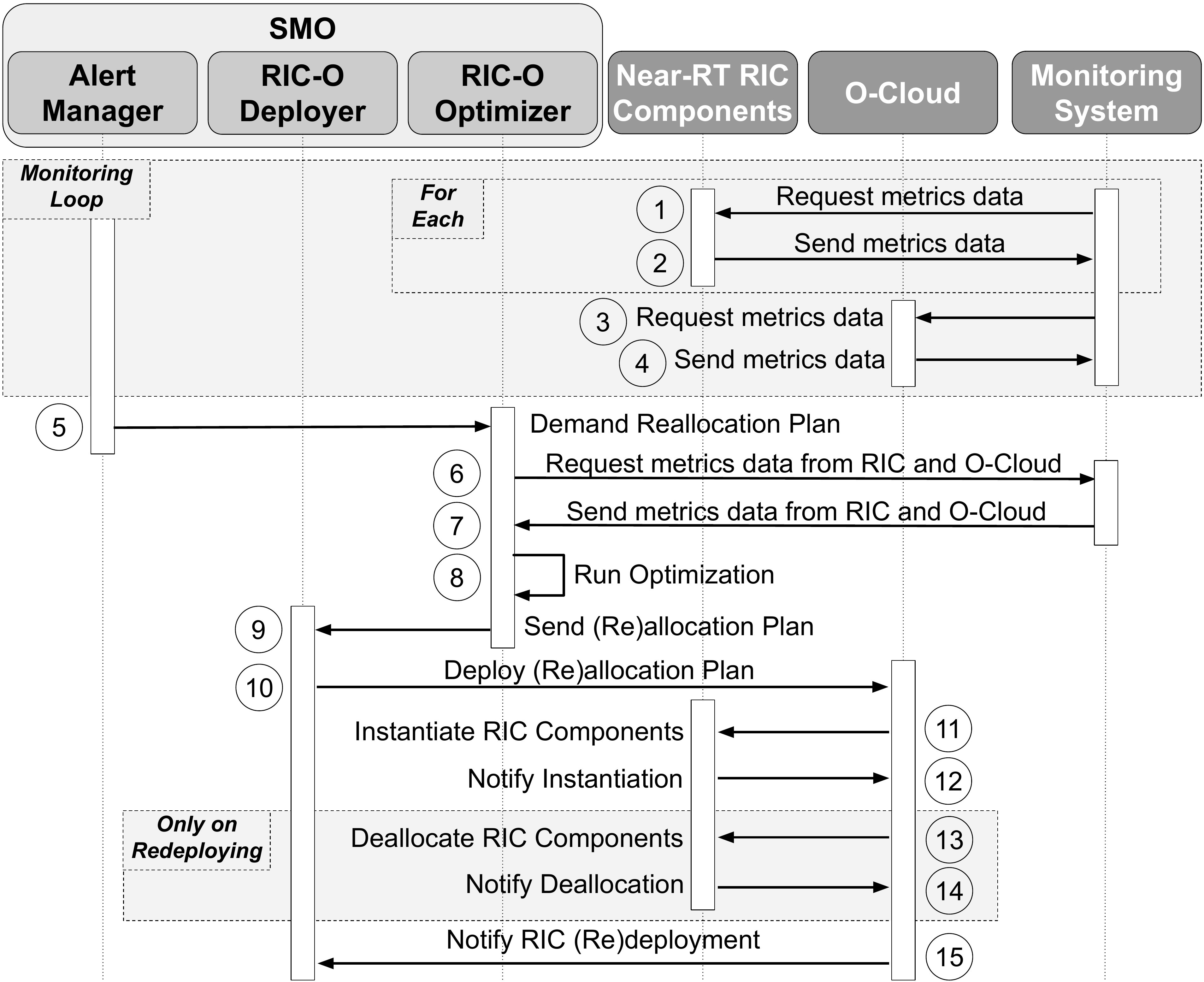}
    \caption{Messages exchanged for Near-RT RIC optimization.}
    \label{fig:sequence}
\end{figure}

Figure~\ref{fig:sequence} shows the message exchange between the RIC-O components, Monitoring System, Near-RT RIC components, and the O-Cloud element while orchestrating the optimization and deployment tasks.
Initially, the deployment is executed as a response to the original request coming from the network operator. Subsequently, a re-deploy is executed only after the \texttt{Alert Manager} sends a notification to the \texttt{RIC-O Optimizer} due to a latency violation in the Near-RT RIC control loop or due to the unavailability of required computing resources in the \texttt{O-Cloud} element.

The \texttt{Monitoring System} shown in Fig.~\ref{fig:sequence} collects metrics regularly from the \texttt{Near-RT RIC Components} and the \texttt{O-Cloud} element. These metrics include CPU and memory usage of Near-RT RIC pods, K8S nodes, and their communication latency (messages 1, 2, 3, and 4). The \texttt{Monitoring Loop} is executed regularly at every second. Once \texttt{Metrics Server} notifies \texttt{Alert Manager} that a given RIC-O trigger has been violated, message 5 is delivered to \texttt{RIC-O Optimizer}.

Afterward, the \texttt{RIC-O Optimizer} starts a reallocation plan to compute the placement of the \texttt{Near-RT RIC Components} that run on the \texttt{O-Cloud} element.
\texttt{RIC-O Optimizer} then requests metrics information that includes the available computing resources of K8S pods, CNs, and the communication latency between them (messages 6 and 7), which is the information required as input for the optimization strategies. Next, the heuristic and optimal strategies are initiated and run in parallel (message 8). Once an outcome from the optimization strategies becomes available, it is sent to the \texttt{RIC-O Deployer} in message 9. Eventually, if there is enough time for computing the optimal solution, two outcomes are generated.
The \texttt{RIC-O Deployer}, in turn, starts deploying the new (re)allocation plan in \texttt{O-Cloud} through the O2 interface defined by O-RAN (message 10).
Near-RT RIC components are instantiated by \texttt{O-Cloud}, which is notified when all the components are deployed (messages 11 and 12). In the particular case when the Near-RT RIC components were deployed previously, \texttt{O-Cloud} releases the resources of the Near-RT RIC components that are no longer required (messages 13 and 14). Finally, \texttt{O-Cloud} notifies the \texttt{RIC-O Deployer} that the deployment has finished (message 15).

\begin{figure}[!h]
    \centering
    \includegraphics[width=0.8\textwidth]{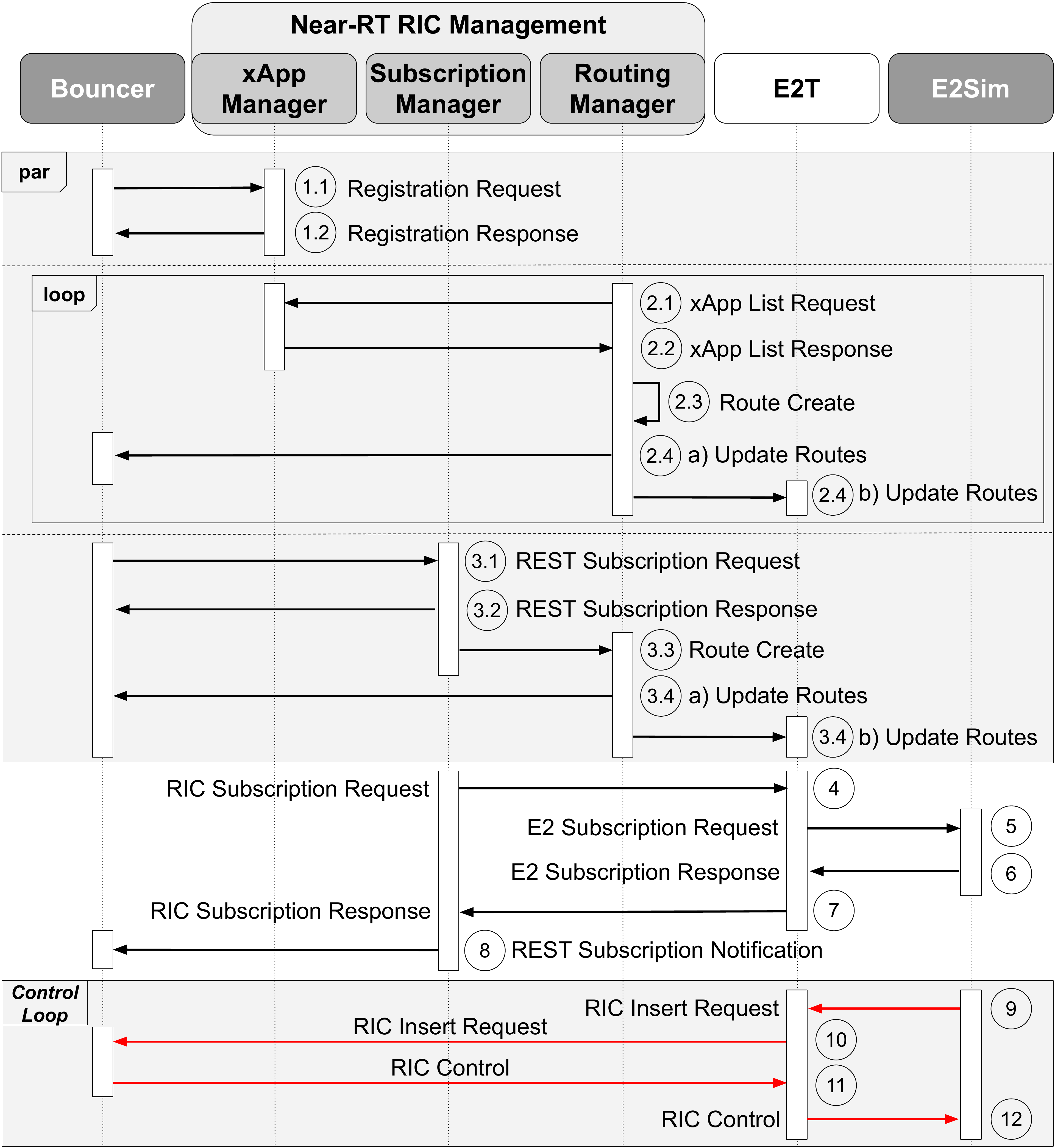}
    \caption{Messages exchanged for xApp deployment and execution.}
    \label{fig:xapp-sequence}
\end{figure}

We also implemented an extended xApp and an E2 simulator (E2Sim), both coded in C++ language.
Our implementation is based on the open-source versions of the Bouncer xApp\footnote{https://gerrit.o-ran-sc.org/r/admin/repos/ric-app/bouncer} and the E2Sim\footnote{https://gerrit.o-ran-sc.org/r/admin/repos/sim/e2-interface} from the O-RAN SC. Moreover, we included new features and updated the service models of both the Bouncer xApp and E2Sim. 
Originally, the Bouncer and E2Sim employed the Key Performance Measurement service model (E2SM-KPM) to expose performance measurements of logical functions running on E2 Nodes \cite{O-RAN.WG3.E2SM-KPM}. However, E2SM-KPM does not support RAN control through the E2AP Control service. Therefore, we implemented the E2SM-RC (RAN Control) \cite{O-RAN.WG3.E2SM-RC} service model, which provides RAN control.
Figure~\ref{fig:xapp-sequence} shows the message exchange between the \texttt{Bouncer} xApp, \texttt{E2Sim}, and Near-RT RIC components. The goal is to monitor the latency-sensitive control loop between the E2 nodes and the corresponding xApp.
We assume that all the Near-RT RIC components and \texttt{E2Sim} are already set up and running when \texttt{Bouncer} xApp is deployed.

When \texttt{Bouncer} xApp initializes, it registers itself with the Near-RT RIC platform by sending a registration message to \texttt{xApp Manager}, which sends back a reply indicating the outcome (messages 1.1 and 1.2). In the meantime, the \texttt{Routing Manager} periodically retrieves the list of xApps registered in \texttt{xApp Manager} to set up the corresponding routing rules in the Near-RT RIC platform (messages 2.1 and 2.2). Moreover, the \texttt{Routing Manager} builds a routing table for all registered xApps (message 2.3) and then distributes it to \texttt{Bouncer} and \texttt{E2T} (messages 2.4a and 2.4b). These routes are required to allow transparent and dynamic message delivery to the endpoints concerned in the communication. During the registration, the \texttt{Bouncer} xApp also subscribes to \texttt{Subscription Manager} to receive messages from the E2AP Insert service from the RAN (messages 3.1 and 3.2). Upon receiving the subscription request, the \texttt{Subscription Manager} notifies \texttt{Routing Manager} to create the corresponding routing rules to deliver those messages to \texttt{Bouncer} xApp (message 3.3). Next, the \texttt{Routing Manager} distributes the generated routing rules to the respective endpoints (messages 3.4a and 3.4b).

Additionally, the \texttt{Subscription Manager} issues a new subscription request (message 4) to the \texttt{E2T}, which implements the E2 interface to communicate with E2 nodes. The \texttt{E2T} then deals with the subscription communication with the \texttt{E2Sim} (messages 5 and 6) according to the subscription that the \texttt{Bouncer} xApp has generated in message 3.1. The communication between \texttt{E2T} and \texttt{E2Sim} is established through Stream Control Transmission Protocol (SCTP). Messages 7 and 8 notify \texttt{Subscription Manager} and \texttt{Bouncer} xApp about the subscription outcome. At this point, \texttt{Bouncer} xApp is fully deployed in the Near-RT RIC platform, and communication with \texttt{E2Sim} can take place through the \texttt{E2T} component.

The \texttt{Control Loop} block in Fig.~\ref{fig:xapp-sequence} illustrates the latency-sensitive control loop of 10ms. When a given E2 node issues a RIC Insert Request message to the Near-RT RIC, the E2 node suspends its execution and waits for the corresponding reply message (i.e., RIC Control) to resume its operation. Messages 9-12 illustrate this control loop that needs to complete within the 10ms threshold. If no RIC Control message is delivered within this threshold time, a default action is executed by the E2 node, which usually implies penalties in the optimization and performance of RAN. Moreover, we implemented the capability of computing the control loop latency between \texttt{E2Sim} and \texttt{Bouncer} xApp. In this case, \texttt{E2Sim} collects the timestamps on sending message 9 and receiving message 12. The latency-sensitive control loop is exported to the Prometheus metrics collector.
\section{Evaluation}\label{sec:eval}


This section presents a performance evaluation of RIC-O both from a theoretical perspective, using analytical modeling (in Subsection~\ref{subsec:eval_sim}), and from a practical perspective, employing real-world experiments (in Subsection~\ref{subsec:eval_real-world}). The first part of the evaluation focuses on scalability, while the second mainly deals with timing issues.

\subsection{Analytical modeling}\label{subsec:eval_sim}

In this part of the evaluation, we assume as RAN topology a next-generation hierarchical network with 512 E2 nodes organized in three main tiers~\cite{morais-placeran:22}. The top one (Tier-1) is closest to the core and so to the cloud node (i.e., $c_0$). The one-way latency (in ms) between the tiers is a uniform random choice from the set $\{1,2,2,3,3\}$, and the one-way latency from Tier-1 to $c_0$ is 4ms. We vary the number of edge computing nodes (i.e., $c_m \in C$) along the network tiers and, for each number of CNs, extract an overlay topology equivalent to the one shown in Fig.~\ref{fig:ran_topo}. There is always a cloud node, and all others are edge computing nodes distributed from bottom to top in the hierarchical RAN network as the number of CNs increases. The cloud node has no fixed cost and presents the following variable cost for Near-RT RIC components: two (RIC\_Man, E2T), one (SDL/STSL, NIBs), and one (xApp). There are two xApps, and both of them access the database, i.e., SDL/STSL and NIBs. Additional information about underlay RAN topology, and Near-RT RIC is presented in Table~\ref{tab:ran_param}.

\begin{table}[ht]
\centering
\begin{tabular}{lccc}
\hline
\rowcolor[HTML]{EEEDEB} 
Parameters            & Tier-1 & Tier-2 & Tier-3 \\ \hline
Number of E2 nodes      & 5 & 20 & 487 \\ \hline
CN fixed cost & 10 & 20 & 30 \\ \hline
\makecell[l]{CN variable cost: (RIC\_Man, E2T), \\(SDL/STSL, NIBs), (xApps)} & 4, 2, 1 & 8, 4, 2 & 16, 8, 4 \\ \hline
$c_m^{Proc}$ (cores)  & 32 & 16 & 8 \\ \hline
$c_m^{Mem}$ (GB) & 64 & 32 & 16 \\ \hline
$c_m^{Sto}$ (GB) & 256 & 256 & 256 \\ \hline
\rowcolor[HTML]{EEEDEB}
Near-RT RIC requirements & $Proc$ & $Mem$ & $Sto$ \\ \hline
RIC\_Man & $4$ & $8$ & $4$ \\ \hline
E2T & $2$ & $4$ & $2$ \\ \hline
SDL/STSL& $2$ & $4$ & $1$ \\ \hline
NIBs& $1$ & $2$ & $50$ \\ \hline
xApps& $1$ & $2$ & $1$ \\ \hline
\end{tabular} 
\caption{Parameters of the underlay RAN topology and requirements of the Near-RT RIC components.}
\label{tab:ran_param}
\end{table}

\begin{figure}[htb]
\centering
    \includegraphics[width=0.7\textwidth]{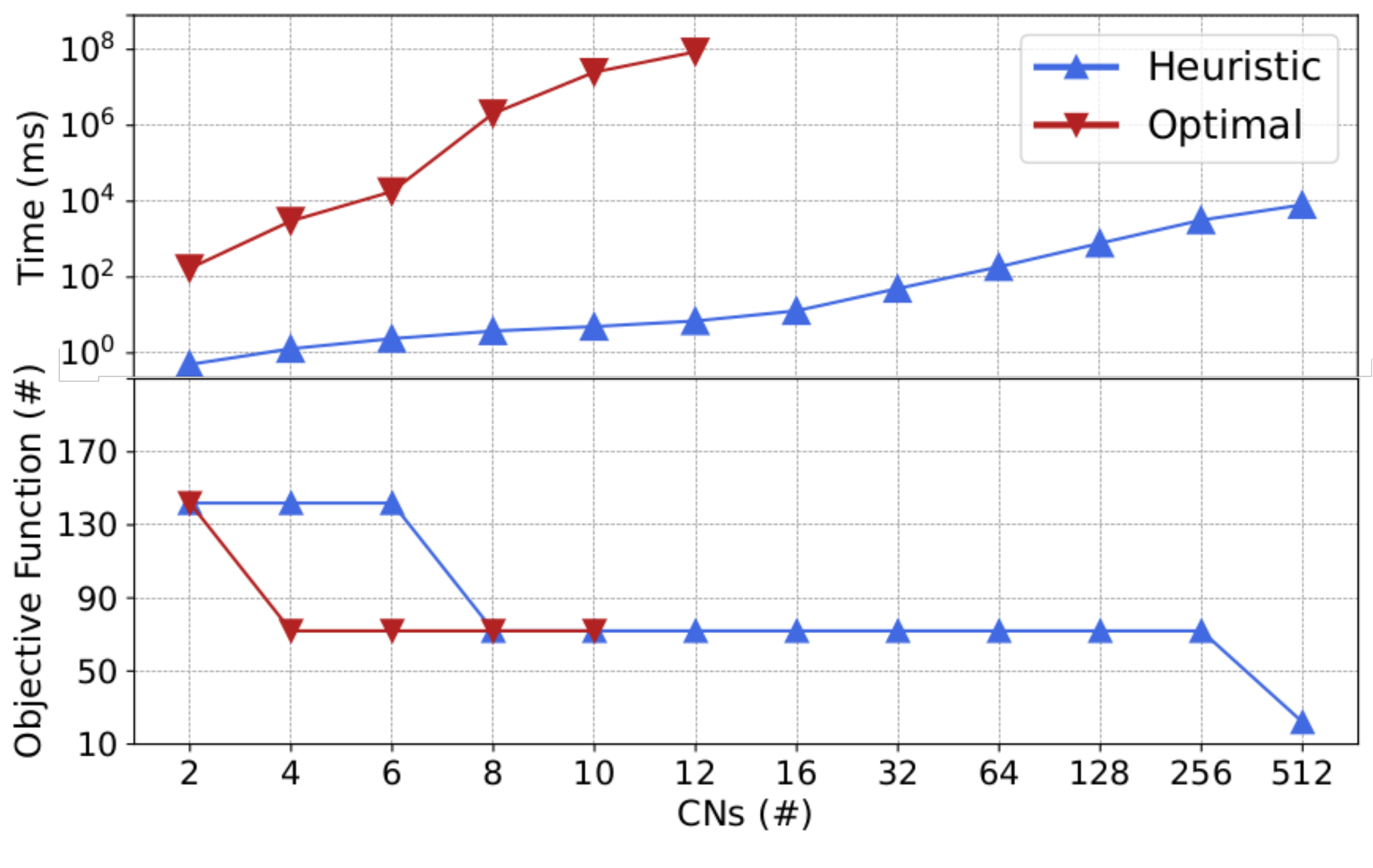}
\caption{Comparison between the optimal and heuristic strategies regarding computing time and total cost (objective function).}
\label{fig:comparing_models}
\end{figure}

We first investigate to which extent the complexity of the exact optimization model (Subsection~\ref{subsec:prob_form}) limits its scalability. Although our heuristic strategy (Subsection~\ref{subseq:heuristic}) is very efficient, it is necessary to investigate the quality of its solutions, i.e., how close to optimal those solutions tend to be. Therefore, we evaluate and compare the optimal and heuristic strategies regarding computing time and objective function as presented in Fig.~\ref{fig:comparing_models}. The computing time is related to the scalability of each strategy, and the objective function, i.e., the total cost, represents the quality of the solution. Moreover, the metrics vary as a function of the number of CNs. 


We can observe in Fig.~\ref{fig:comparing_models} that the optimal strategy scales poorly, being unable to find a solution for a RAN with 12 CNs even after running for an entire day. Therefore, the optimal strategy is rarely helpful for large real-world instances. Nevertheless, the heuristic strategy scales very well and finds a solution for 512 CNs in less than 10 seconds. Concerning the quality of the solutions, as expected, the heuristic strategy cannot always find the optimal solution. However, a decreasing trend can be discerned in the objective function that suggests the heuristic strategy is in the right direction, including in scenarios with a large number of CNs, e.g., 512 CNs. In those cases, the cheapest edge computing nodes (i.e., those in Tier-1) are readily available, and the solution improves noticeably.

\begin{figure}[htb]
\centering
    \includegraphics[width=0.7\textwidth]{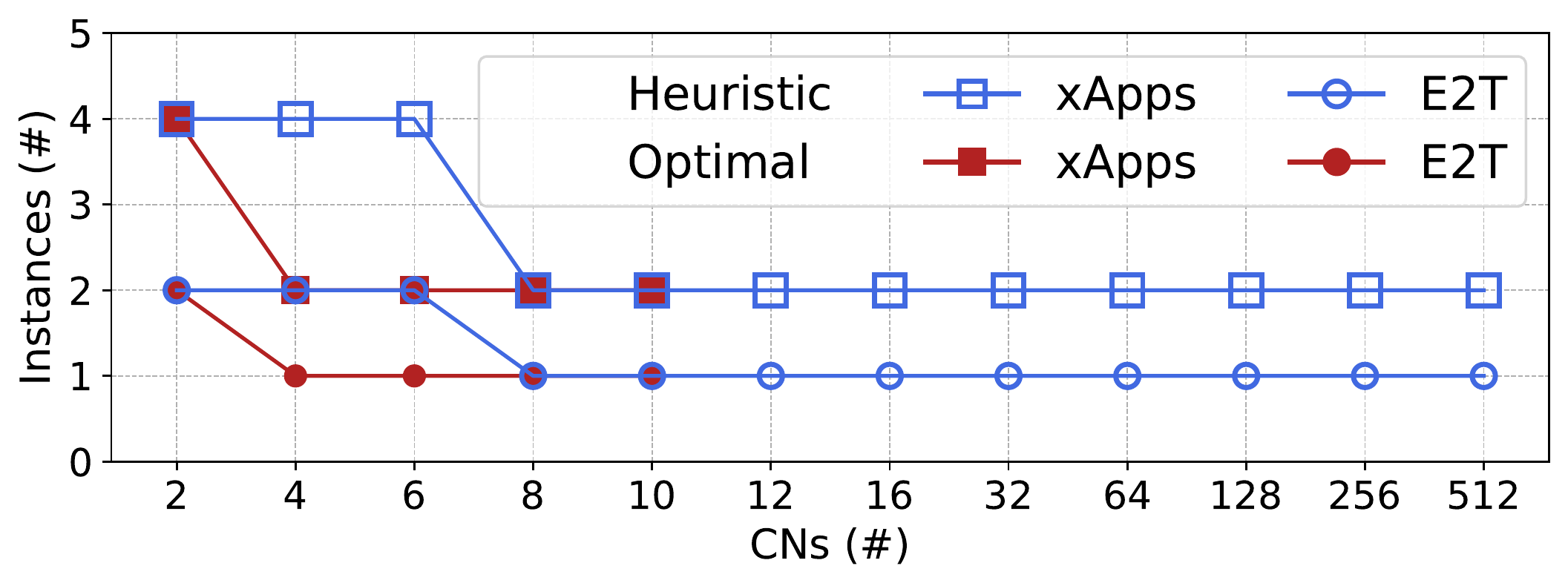}
\caption{Comparison between optimal and heuristic strategies regarding the number of xApps and E2T instances.}
\label{fig:instances}
\end{figure}

We further investigate the optimization strategies by evaluating the number of instances of the main Near-RT RIC components involved in the latency-sensitive control loop, i.e., xApps and E2T. As presented in Fig.~\ref{fig:instances}, our heuristic strategy finds the best (i.e., the smallest) number of E2T instances, matching the result obtained by the optimal strategy when this is available and keeping it at the minimum (i.e., one instance) as the number of CNs increases. Moreover, the heuristic strategy finds some non-optimal solutions for the number of xApps instances, e.g., with four and six CNs. This behavior is expected from our heuristic strategy because it prioritizes Near-RT RIC components involved in the latency-sensitive control loop in the following order: E2T, xApps, SDL/STSL, and NIBs. Since latency optimization and cost optimization are performed in this order and without revisiting previous decisions for improvement, the heuristic strategy potentially finds non-optimal solutions involving xApps, SDL/STSL, or NIBs.

\subsection{Real-world experiments}\label{subsec:eval_real-world}

In this part of the evaluation, we run the experiments in a scenario with five CNs, which are virtual machines (VMs) with the following configuration: 4 vCPUs, 8 GB RAM, and 50 GB of the virtual disk. One CN represents the cloud node (i.e., $c_0$), and the others represent the edge computing nodes (i.e., $c_m \in C$). These CNs are worker nodes in a K8S cluster managed by a master node running a sixth VM with the following configuration: eight vCPUs, 16 GB RAM, and 100 GB of the virtual disk. All VMs are hosted on a DELL PowerEdge M610 server with four Intel Xeon X5660 processors and 192 GB RAM, which runs VMware ESXi 6.7 as the hypervisor. Additional details about the software tools employed in the experiments are available in the public repository of this article. We also employed an E2 simulator to represent four E2 nodes that must be served by the Near-RT RIC. Overlay and underlay topologies are the same in this part of the evaluation and correspond to the one presented in Fig.~\ref{fig:ran_topo}.


To illustrate the orchestration capabilities of RIC-O, we designed two scenarios in which the latency-sensitive control loop is disrupted and show how our proposal acts to bring the Near-RT RIC back to normal operation. In the first scenario, RIC-O must deal with a sudden and high increase in the latency of the path used to serve a certain E2 node. The second scenario is more challenging because RIC-O needs to deal with a CN that becomes unavailable, i.e., any latency-sensitive control loop involving this CN disappears since the Near-RT RIC components running in it suddenly become inaccessible.

\begin{figure}[!h]
\centering
    \includegraphics[width=0.75\textwidth]{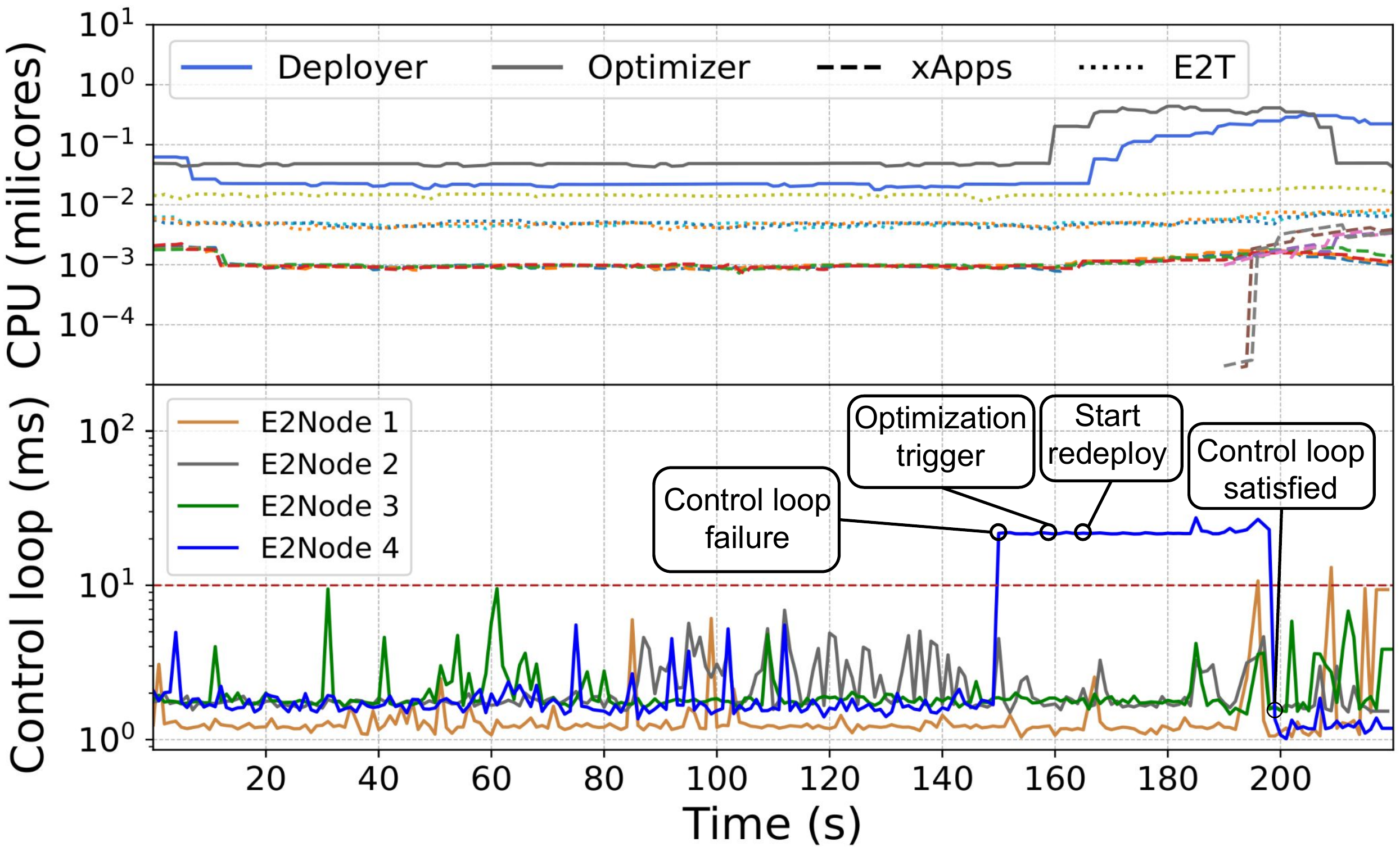}
\caption{Reaction to a sudden violation of requirements of the latency-sensitive control loop.}
\label{fig:control_loop}
\end{figure}

In a real-world RAN, the latency between a pair of nodes may change due to a (re)route decision in the underlying network, for example. Since our underlay network matches the overlay one, we emulate the sudden increase in the latency between the E2 node and its corresponding E2T by reconfiguring the latency in the virtual link connecting these nodes. Figure~\ref{fig:control_loop} illustrates the main events occurring along the time in this scenario. This figure shows the status of the control loop between each E2 node and its corresponding xApp. In addition, the figure presents the CPU utilization of some essential software components (i.e., \texttt{RIC-O Deployer}, \texttt{RIC-O Optimizer}, xApps, and \texttt{E2T}), which helps keep track of the actions performed by RIC-O.


As illustrated in Fig.~\ref{fig:control_loop}, the first scenario is initially in a fully operational state, and the latency-sensitive control loop of each E2 node is satisfied by the Near-RT RIC thanks to the initial orchestration defined by RIC-O. Then, at time instant 150s, the latency of the control loop from E2 node 4 increases suddenly and remains persistently above 10ms, as indicated by event \textit{Control loop failure} in the figure. After 10 seconds, the \texttt{Monitoring System} considers that the event is a consistent control loop violation and notifies the \texttt{RIC-O Optimizer} to compute a new solution, at time instant 160s, as indicated by the event \textit{Optimization trigger}. The heuristic strategy of the \texttt{RIC-O Optimizer} quickly finds a solution and requests \texttt{RIC-O Deployer} to apply this new placement nearly 5 seconds later, as indicated by the event \textit{Start redeploy}. The \texttt{RIC-O Optimizer} keeps running the optimal strategy thread. Finally, the redeploy of the Near-RT RIC components and reconfiguration of E2 nodes completes at time instant 200s, as indicated by event \textit{Control loop satisfied}, when the latency-sensitive control loop is again limited to 10ms.

\begin{figure}[!h]
\centering
    \includegraphics[width=0.7\textwidth]{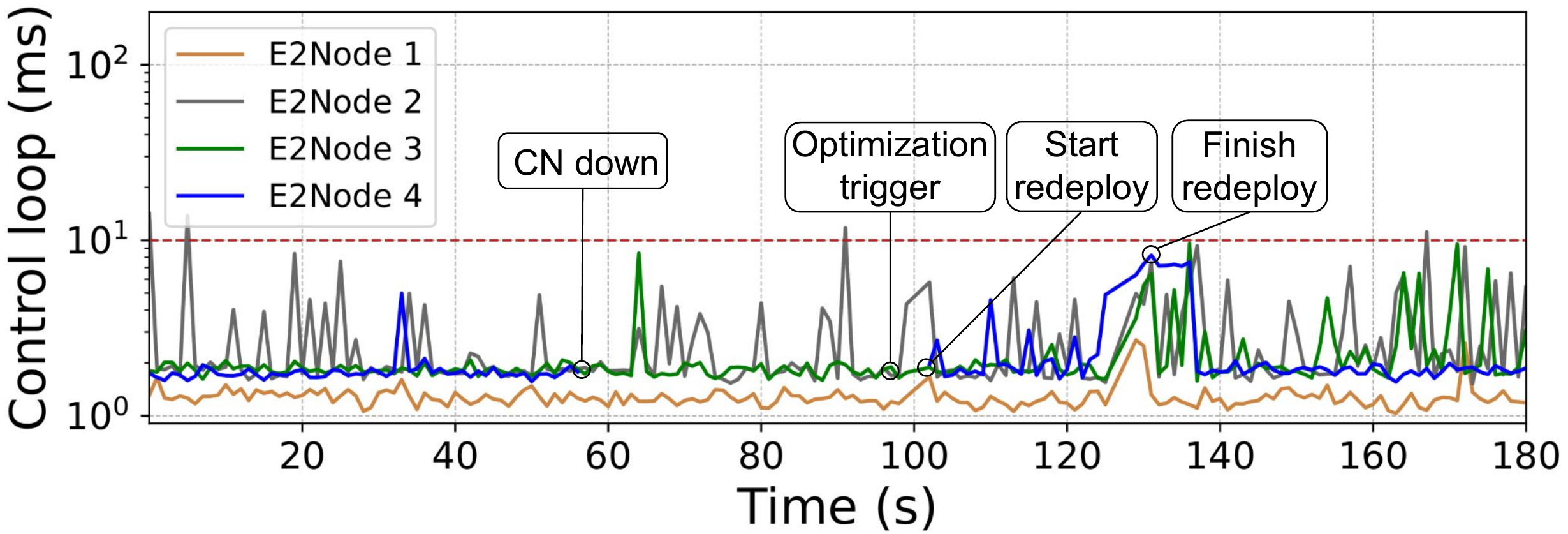}
\caption{Reaction to sudden unavailability of CN under use.}
\label{fig:CN_down}
\end{figure}

Figure~\ref{fig:control_loop}, and also Fig.~\ref{fig:CN_down} show a few measurements of the control loop that go above 10ms. This behavior is related to the underlying operating system and virtualization platform (i.e., hypervisor). A tight threshold of 10ms is on such a sensitive scale that even a traditional process scheduler may sometimes cause a small variation. Since fine-tuning those systems are out of the scope of this work, we configured the \texttt{Monitoring System} to report only persistent violations of the 10ms threshold in latency-sensitive control loops. 


The second scenario, in which a CN suddenly crashes, may either represent a software or hardware failure, or network outage. We emulate this behavior by abruptly forcing a shutdown of the VM running the CN. Figure~\ref{fig:CN_down} illustrates the main events occurring along the time in this scenario. Similar to Fig.~\ref{fig:control_loop}, Fig.~\ref{fig:CN_down} also shows the status of the control loop between each E2 node and its corresponding xApp. However, we have not identified relevant information that justified presenting measurements related to Near-RT RIC nor RIC-O components.  

As illustrated in Fig.~\ref{fig:CN_down}, the second scenario also starts from a fully operational state, where the latency-sensitive control loops of all E2 nodes are satisfied by the Near-RT RIC thanks to the initial orchestration defined by RIC-O. The CN running Near-RT RIC components responsible for serving the E2 node 4 suddenly becomes unavailable, as indicated by the event \textit{CN down} in the figure. In this case, the latency-sensitive control loop of this E2 node is totally disrupted, i.e., there are no more measurements of the control loop. After 50 seconds, the \texttt{Monitoring System} detects the problem and notifies the \texttt{RIC-O Optimizer} to compute a new solution, as indicated by the event \textit{Optimization trigger}. This time interval for reporting the problem may seem long, but it is the default K8S policy for detecting worker node unavailability. Again, after nearly 5 seconds, the heuristic strategy of \texttt{RIC-O Optimizer} finds a solution and requests \texttt{RIC-O Deployer} to apply the new placement, as indicated by the event \textit{Start redeploy}. Finally, the redeployment of the Near-RT RIC components finishes, and the latency-sensitive control loop of E2 node 4 is reestablished at time instant 130s, as indicated by the event \textit{Finish redeploy}.

\section{Conclusion}\label{sec:conclusion}

In this work, we claim that the most efficient approach to ensure that the Near-RT RIC is able to deal with a large real-world RAN is by disaggregating and distributing its components. However, it is necessary to orchestrate these components properly across the edge-cloud continuum to ensure the latency-sensitive control loop that involves xApps running over the Near-RT RIC. Therefore, we propose the RIC orchestrator (RIC-O) that employs a hybrid strategy combining optimal and heuristic strategies for efficient placement of the Near-RT RIC components. We evaluate RIC-O through analytical modeling and real-world experiments to illustrate its properties and benefits. 
Future work includes the investigation of machine learning-based time series forecasting for improving the orchestration of Near-RT RIC components.
Moreover, the adoption of the O-RAN architecture by large-scale production networks will provide additional data on its usage, thus allowing extensive investigation of novel optimization techniques. 
Finally, we envision the potential of the RIC-O being adopted as part of the Non-RT RIC, running as an rApp, which will demand further experiments and software integration.

\section*{Acknowledgment}

This work was conducted with partial financial support from the National Council for Scientific and Technological Development (CNPq) under Grant Nos. 405111/2021-5 and 130555/2019-3 and from the Coordination for the Improvement of Higher Education Personnel (CAPES) - Finance Code 001, Brazil. Moreover, RNP partially supported the work, with resources from MCTIC, Grant No. 01245.010604/2020-14, under the 6G Mobile Communications Systems project and by MCTIC/CGI.br/São Paulo Research Foundation (FAPESP) through the Project SAMURAI - Smart 5G Core And MUltiRAn Integration under Grant 2020/05127-2 and Programmability, ORchestration and VIRtualization of 5G Networks (PORVIR-5G) under Grant No. 2020/05182-3.


\bibliographystyle{IEEEtran}

\begin{thebibliography}{10}
\providecommand{\url}[1]{#1}
\csname url@samestyle\endcsname
\providecommand{\newblock}{\relax}
\providecommand{\bibinfo}[2]{#2}
\providecommand{\BIBentrySTDinterwordspacing}{\spaceskip=0pt\relax}
\providecommand{\BIBentryALTinterwordstretchfactor}{4}
\providecommand{\BIBentryALTinterwordspacing}{\spaceskip=\fontdimen2\font plus
\BIBentryALTinterwordstretchfactor\fontdimen3\font minus
  \fontdimen4\font\relax}
\providecommand{\BIBforeignlanguage}[2]{{%
\expandafter\ifx\csname l@#1\endcsname\relax
\typeout{** WARNING: IEEEtran.bst: No hyphenation pattern has been}%
\typeout{** loaded for the language `#1'. Using the pattern for}%
\typeout{** the default language instead.}%
\else
\language=\csname l@#1\endcsname
\fi
#2}}
\providecommand{\BIBdecl}{\relax}
\BIBdecl

\bibitem{pana22-5g-survey}
\BIBentryALTinterwordspacing
V.~S. Pana, O.~P. Babalola, and V.~Balyan, ``{5G radio access networks: A
  survey},'' \emph{Array}, vol.~14, p. 100170, 2022. [Online]. Available:
  \url{https://www.sciencedirect.com/science/article/pii/S2590005622000315}
\BIBentrySTDinterwordspacing

\bibitem{O-RAN-Alliance}
\BIBentryALTinterwordspacing
{O-RAN Alliance}, ``{O-RAN Alliance},'' 2022. [Online]. Available:
  \url{https://www.o-ran.org/}
\BIBentrySTDinterwordspacing

\bibitem{garcia-saavedra2021-oran}
A.~Garcia-Saavedra and X.~Costa-Pérez, ``{O-RAN: Disrupting the Virtualized
  RAN Ecosystem},'' \emph{IEEE Communications Standards Magazine}, vol.~5,
  no.~4, pp. 96--103, 2021.

\bibitem{oran-arch}
{O-RAN Alliance}, ``{O-RAN Architecture Description},'' {O-RAN Alliance}, Tech.
  Rep. O-RAN.WG1.O-RAN-Architecture-Description-v07.00, 2022.

\bibitem{3gpp-ts-38.401}
3GPP, ``{3GPP TS 38.401 V17.1.1 - NG-RAN; Architecture description},'' {3rd
  Generation Partnership Project (3GPP)}, Tech. Rep., 2022.

\bibitem{balasubramanian2021-ric}
B.~Balasubramanian \emph{et~al.}, ``{RIC: A RAN Intelligent Controller Platform
  for AI-Enabled Cellular Networks},'' \emph{IEEE Internet Computing}, vol.~25,
  no.~2, pp. 7--17, 2021.

\bibitem{rimedolabs2022-near-rt-ric-arch}
M.~Dryjański and A.~Kliks, ``{The O-RAN Whitepaper 2022 RAN Intelligent
  Controller, xApps and rApps},'' RIMEDO Labs, Tech. Rep., 2022.

\bibitem{Salvo22}
S.~D’Oro \emph{et~al.}, ``{OrchestRAN: Network Automation through
  Orchestrated Intelligence in the Open RAN},'' in \emph{IEEE Conference on
  Computer Communications (INFOCOM)}, 2022, pp. 270--279.

\bibitem{O-RAN.WG3.E2GAP}
{O-RAN Alliance}, ``{Near-Real-time RAN Intelligent Controller Architecture \&
  E2 General Aspects and Principles},'' {O-RAN Alliance}, Tech. Rep.
  O-RAN.WG3.E2GAP-v02.02, 2022.

\bibitem{O-RAN.WG3.E2AP}
------, ``{Near-Real-time RAN Intelligent Controller, E2 Application Protocol
  (E2AP)},'' {O-RAN Alliance}, Tech. Rep. O-RAN.WG3.E2AP-v02.03, 2022.

\bibitem{oran-use-cases}
------, ``{O-RAN.WG1.Use-Cases-Detailed-Specification-v09.00},'' {O-RAN
  Alliance}, Tech. Rep., 2022.

\bibitem{oran-sc}
\BIBentryALTinterwordspacing
------. {O-RAN Software Community (SC)}. [Online]. Available:
  \url{\url{https://oran-osc.github.io}}
\BIBentrySTDinterwordspacing

\bibitem{Kumar22}
A.~K. Singh and K.~Khoa~Nguyen, ``{Joint Selection of Local Trainers and
  Resource Allocation for Federated Learning in Open RAN Intelligent
  Controllers},'' in \emph{IEEE Wireless Communications and Networking
  Conference (WCNC)}, 2022, pp. 1874--1879.

\bibitem{Huff21}
A.~Huff, M.~Hiltunen, and E.~P. Duarte, ``{RFT: Scalable and Fault-Tolerant
  Microservices for the O-RAN Control Plane},'' in \emph{IFIP/IEEE
  International Symposium on Integrated Network Management (IM)}, 2021, pp.
  402--409.

\bibitem{Schmidt21}
R.~Schmidt, M.~Irazabal, and N.~Nikaein, ``{FlexRIC: an SDK for next-generation
  SD-RANs},'' in \emph{17th International Conference on emerging Networking
  EXperiments and Technologies (CoNEXT)}, 2021, p. 411–425.

\bibitem{Bharath21}
B.~Balasubramanian \emph{et~al.}, ``{RIC: A RAN Intelligent Controller Platform
  for AI-Enabled Cellular Networks},'' \emph{IEEE Internet Computing}, vol.~25,
  no.~2, pp. 7--17, 2021.

\bibitem{Cao21}
Y.~Cao \emph{et~al.}, ``{Federated Deep Reinforcement Learning for User Access
  Control in Open Radio Access Networks},'' in \emph{IEEE International
  Conference on Communications}, 2021, pp. 1--6.

\bibitem{Cao22}
------, ``{User Access Control in Open Radio Access Networks: A Federated Deep
  Reinforcement Learning Approach},'' \emph{IEEE Transactions on Wireless
  Communications}, vol.~21, no.~6, pp. 3721--3736, 2022.

\bibitem{Johnson21}
D.~Johnson, D.~Maas, and J.~V.~D. Merwe, ``{NexRAN: Closed-loop RAN slicing in
  POWDER -A top-to-bottom open-source open-RAN use case},'' in
  \emph{Proceedings of the 15th ACM Workshop on Wireless Network Testbeds,
  Experimental evaluation (WiNTECH)}, 2021, pp. 17--23.

\bibitem{MIQP_complexity}
A.~D. Pia \emph{et~al.}, ``{Mixed-integer quadratic programming is in NP},''
  \emph{Mathematical Programming}, vol. 162, no.~1, pp. 225--240, 2017.

\bibitem{ric-alarm20}
\BIBentryALTinterwordspacing
{O-RAN SC}, ``{RIC Alarm System},'' 2020. [Online]. Available:
  \url{https://wiki.o-ran-sc.org/display/RICP/RIC+Alarm+System}
\BIBentrySTDinterwordspacing

\bibitem{O-RAN.WG3.E2SM-KPM}
{O-RAN Alliance}, ``{Near-Real-time RAN Intelligent Controller E2 Service Model
  (E2SM) KPM},'' {O-RAN Alliance}, Tech. Rep. O-RAN.WG3.E2SM-KPM-v02.02, 2022.

\bibitem{O-RAN.WG3.E2SM-RC}
------, ``{Near-Real-time RAN Intelligent Controller E2 Service Model (E2SM),
  RAN Control},'' {O-RAN Alliance}, Tech. Rep. ORAN.WG3.E2SM-RC-v01.02, 2022.

\bibitem{morais-placeran:22}
F.~Z. Morais, G.~M.~F. De~Almeida, L.~L. Pinto, K.~Cardoso, L.~M. Contreras,
  R.~d.~R. Righi, and C.~B. Both, ``{PlaceRAN: optimal placement of virtualized
  network functions in Beyond 5G radio access networks},'' \emph{IEEE Trans.
  Mobile Computing}, pp. 1--1, 2022.

\end{thebibliography}


\end{document}